%% file: main.tex
\documentclass[acmsmall,screen,nonacm]{acmart}
\input{includes}

\newsavebox{\plotbox}\newsavebox{\tabcapbox}
\usepackage{float}
\usepackage{wrapfig}
\begin{document}
\sloppy

\title{A View from the Bridge: Detecting Boundary Type Errors
in Native Extensions of Dynamic Languages}

\authorsaddresses{}
\author{%
\NoCaseChange{\textnormal{\makebox[\textwidth][s]{\Large Thodoris Sotiropoulos\textsuperscript{1,2,*}\hfill Georgios Alexopoulos\textsuperscript{1,2,*}\hfill Dimitris Mitropoulos\textsuperscript{1,2}}}\\[6pt]
\textnormal{\makebox[\textwidth][c]{\normalsize \textsuperscript{1}University of Athens, Greece}}\\[1pt]
\textnormal{\makebox[\textwidth][c]{\normalsize \textsuperscript{2}National Infrastructures for Research and Technology (GRNET), Greece}}\\[1pt]
\textnormal{\makebox[\textwidth][c]{\normalsize \{theosotir, grgalex, dimitro\}@ba.uoa.gr}}}%
}

\begin{abstract}
Programs written in dynamic languages,
such as Python and JavaScript,
often rely on native extensions written in C, C++ or Rust
for performance or access to system libraries.
Unlike the host language,
native code is not safe.
It accesses the raw memory of the values
it receives from the host,
so before using a value,
it must check the value's \emph{tag},
which determines its memory layout.
When this check is missing,
or accepts tags incompatible with the one
that the native code later assumes,
a \emph{boundary type error} occurs,
and the program crashes or
silently corrupts its data.
Our key insight is
that finding these errors requires
no inference of host-language types.
Native code must check
every value it obtains from the host,
so the code itself provides all the information we need:
its checks indicate which tags a value may have,
while its operations reveal which tag they assume.
This differs from prior work on statically typed languages,
which infers \emph{host-language} types from native code
and checks them against the parameter types
that the host program declares for each native function.
Dynamic languages lack such declarations.

We capture what native code does with host values in XLIR,
a small language in which we define boundary type errors.
We then detect them
with a flow-sensitive analysis
that tracks the tags each check establishes.
At each operation,
we encode whether some value can
pass the preceding checks
yet violate the tag that the operation assumes
as a satisfiability problem,
which we solve with an SMT solver.
We have implemented our approach
in a tool called \tool,
which analyzes the LLVM IR of Node.js
and CPython extensions.
It detects all
\empirical{72} known errors in an
existing JavaScript benchmark
and \empirical{33} of
\empirical{42} reported in
fixed GitHub issues
of popular PyPI packages.
In the wild,
\tool\ finds thousands of previously unknown type errors
in the native code of \empirical{268} npm
and \empirical{204} PyPI packages,
including popular ones such as
{\tt Pillow}, {\tt wrapt} and {\tt canvas},
and in the built-in standard library extensions of CPython and Node.js.
It analyzes most extensions in under ten seconds,
and the errors it finds range from hard crashes
to silently incorrect output.
We have filed \empirical{58} issues
in \empirical{45} of these projects,
including CPython and Node.js,
of which \empirical{26} have been fixed.
\end{abstract}



\renewcommand{\shortauthors}{Sotiropoulos, Alexopoulos, and Mitropoulos}
\maketitle
\begingroup\renewcommand{\thefootnote}{$*$}%
\footnotetext{Both authors contributed equally.}\endgroup

\section{Introduction}
\label{sec:intro}

Modern software is rarely written in a single
language~\cite{RNHKM24,charon,pyxray,gasket}.
Prominent examples are JavaScript and Python programs
running on Node.js and CPython,
which often rely on \emph{native extensions},
i.e., compiled modules written in
C, C++ or Rust~\cite{python-capi,node-api}.
Native extensions speed up performance-critical code
and expose functionality
that the host language lacks,
such as file system and network
access~\cite{brown-checker,pyxray,gasket}.
They also enable developers to leverage
existing third-party native
libraries~\cite{crosseco,insight}.
As a result,
native extensions are widespread.
On PyPI,
\empirical{59\%} of the top 100,000 packages
have at least one dependency
that ships a native extension~\cite{crosseco},
and even CPython and Node.js implement
much of their own standard libraries
as native extensions~\cite{pystdlib,brown-checker,gasket}.

On the one hand,
dynamic languages are~\emph{safe}:
their runtime, the \emph{host},
marks each value with a \emph{tag}
that determines its layout in memory,
and checks this tag before every operation
so that no operation misreads the memory
of a value~\cite{Cardelli97,brown-checker}.
On the other hand,
native code gets no such protection.
It interacts with the host
through the host's \emph{binding API},
e.g., N-API~\cite{node-api} for Node.js
and the C-API~\cite{python-capi} for CPython.
It receives every host value
as a handle of one~\emph{general} type,
such as \code{PyObject *},
and must itself bring the value
into a statically typed native representation,
such as a C struct or an instance of a C++ class,
before it can work with it.
Brown et al.~\cite{brown-checker} call
this step a translation.
We call it a \emph{use},
since everything that native code does
with the value afterwards builds on it.
A use accesses the memory of the value
according to one fixed layout,
and thus \emph{assumes}
that the value has the tag that determines that layout
(e.g., reading the length of an array
assumes the tag \code{Array}).
Neither the host nor the compiler
checks this assumption.
A use is safe only
if native code first establishes the assumed tag
with a \emph{tag check},
which the binding API provides.

\point{The problem}
When a host value does not have
the tag that native code assumes,
the use interprets its memory
with the wrong layout.
It may then follow a bogus pointer
and crash the process,
or read the wrong bytes
and silently corrupt
the program's data~\cite{brown-checker,flowjs,YC25}.
We call such an unsafe use
a \emph{boundary type error}.
Prior work attributes these errors
to missing checks~\cite{brown-checker,flowjs}.
We find that many errors occur
even though native code performs a check
(\cref{sec:rq3-bug-characteristics}),
because the check is too weak:
it also accepts values that lack the tag the use assumes,
or it does not examine the tag at all.
Our approach uncovered a case of the latter
in CPython's own \code{\_io} module
(\cref{fig:motivating-b}).
To write text to a file,
the native function \code{write}
first calls an encoder,
which a Python program can define,
to convert the text into the desired sequence of bytes.
When the encoder returns the text unchanged,
\code{write} assumes that the text is ASCII,
without checking its tag.
Hence,
writing \code{"DELETE: NO\texttrademark"}
through such an encoder
stores only \code{DELETE} in the file,
leading to data loss.

\point{Existing work}
Existing approaches miss many
of these errors.
Prior static analyses of binding
code in dynamic languages~\cite{brown-checker,flowjs}
have two limitations.
First,
they look only at the arguments of a native function,
and do not consider the values
that native code obtains from the host afterwards,
such as the result of a callback.
Developers tend to neglect these values too:
they account for~\empirical{51\%}
of the errors we find,
and in~\empirical{84\%} of those cases
the code performs no check on such values
(\cref{sec:rq3-bug-characteristics}).
Second,
they treat \emph{any} check
as a \emph{complete} check.
For example,
in~\cref{fig:motivating-a},
native code checks
only that a value is an object,
but then uses it as an instance of the class \code{Context2d},
i.e., assumes the tag \code{Context2d}.
Native code then reads
its memory as if it were a \code{Context2d},
obtains an invalid pointer,
and crashes the process.
Yet prior work considers such a use safe.
Such incomplete checks account for 38\%
of the errors in arguments that we find
(\cref{sec:rq3-bug-characteristics}).

Furr and Foster~\cite{FF05,FF08}
check native extensions of statically typed languages,
namely OCaml and Java.
There, the host program declares the type
of every value that it passes to native code,
e.g., the Java method \code{native int f(String s)}
guarantees that \code{s} is a string.
Their checkers infer the~\emph{host-language} type
that native code assumes for each value
and check it against the declared one.
Dynamic languages declare no such types.
Inferring them from the scripts that use an extension
would require record types for objects
and higher-order types for callbacks,
which is hard for dynamic languages,
and impossible when no such script is available.
%
\input{figures/motivating}

\point{Approach}
Our key observation is that,
at the level of tags,
native code contains all the information needed
to detect these errors,
\emph{without} inferring host-language types, such as record or function types.
Native code knows nothing about a host value
until it tests it.
Its~\emph{tests},
i.e., checks of a value's tag,
establish the value's layout,
and its~\emph{uses} state which layout they assume.
Beyond its arguments,
native code can obtain values from the host,
e.g., by reading an object member or calling a callback.
Each such request is a~\emph{host operation}:
control passes to the host,
which may run arbitrary script code
before returning a result.
The result can therefore have any tag,
and native code must test it before using it,
regardless of what earlier tests established.
Detecting boundary type errors thus reduces to
\textit{checking that every use is preceded by a test
that establishes the tag the use assumes}.
We capture this in XLIR,
a small language whose values are tags
and whose only operations on them
are tests, uses and host operations.

XLIR brings the problem into a form
in which we can reason about boundary type errors efficiently.
Our approach transforms native extensions into XLIR programs
by leveraging a specification of the binding API,
which states what each of its functions does with host values
(e.g., the tag it tests or assumes).
To reason about the possible tags of each value
at each program point,
we introduce \emph{tag refinement},
a flow-sensitive analysis in the tradition of
occurrence typing~\cite{THF08,THF10}
and flow typing~\cite{GSK11,Pearce13}.
Each test on a value refines its possible tags.
Finally,
an SMT-based subtyping check~\cite{BGHL12}
decides,
for each use,
whether every possible tag of the value
is compatible with the tag the use assumes.
If not,
we consider the use unsafe,
and report a concrete value
that passes all preceding tests (if any),
yet violates the assumption of native code.

\point{Results}
We implement our approach in a tool called \tool,
which analyzes native extensions
of CPython and Node.js.
\tool\ detects all \empirical{72} known errors
in an existing JavaScript benchmark~\cite{flowjs}
and \empirical{33} of the \empirical{42} errors
reported in fixed GitHub issues of popular PyPI packages.
In the wild,
it finds thousands of previously unknown errors
in \empirical{268} npm and \empirical{204} PyPI packages,
as well as in the built-in extensions of CPython and Node.js.
More than half of them crash the process,
and many others silently corrupt its output or state.
\tool\ is also fast:
it analyzes an extension in under a second at the median,
and most extensions in under ten seconds.
We have filed \empirical{58} issues
in \empirical{45} of these projects,
of which \empirical{26} have been fixed.

\point{Contributions}
In summary,
we make the following contributions:
\begin{itemize}
\item We reduce the detection of boundary type errors
in dynamic languages
to checking \emph{tag safety},
and capture this in XLIR,
a small language of tags
(\cref{sec:approach}).
\item We present \tool,
which checks tag safety
for native extensions of CPython and Node.js.
It lowers the LLVM IR of an extension into XLIR
using a specification of the binding API,
and combines \emph{tag refinement},
a flow-sensitive analysis of the possible tags of each value,
with an SMT-based subtyping check
that produces a witness value for each detected type error
(\cref{sec:practical}).
\item We evaluate \tool\ on known errors
and in the wild,
on npm and PyPI packages
and on the built-in extensions of CPython and Node.js
(\cref{sec:evaluation}).
Using \tool,
we have filed \empirical{58} issues
in \empirical{45} projects,
including CPython and Node.js,
of which \empirical{26} have been fixed.
\end{itemize}

\section{Background and Motivation}
\label{sec:background}

\subsection{The Host--Native Boundary in Dynamic Languages}
\label{sec:bg-apis}\label{sec:bg-safety}\label{fig:layout}\label{fig:layout-v8}\label{sec:prelims}

\input{figures/fig-interaction}

\point{Host}
The \emph{host} is the runtime of a dynamically typed language,
such as V8 for JavaScript and CPython for Python.
It executes a \emph{script},
such as the JavaScript program
at the top right of Figure~\ref{fig:interaction}.
When the script creates a value,
such as the array \code{[3, 4]},
the host allocates it in memory
and marks it with a~\emph{tag},
which determines its~\emph{layout},
i.e., how the value is arranged in memory.
For example,
the layout of the tag \code{Array} stores
the length and the elements of the array
(the ``raw bits'' in the figure).
A tag is~\emph{not} the value's dynamic type in the language.
For example,
CPython lays out ASCII strings
differently from other strings
and gives them a different tag,
although both have the Python type~\code{str}.
A script can also change the members of a JavaScript object
and even its prototype,
which decides what \code{instanceof} reports,
but never its tag.
Tags form a~\textit{hierarchy}.
A tag $t$ lies \emph{below} a tag $t'$
when the layout of $t$ is compatible with that of $t'$.
For example,
\code{Array} lies below \code{Object}:
an \code{Array} holds
all the fields of an \code{Object}
at the same positions,
so code that reads an \code{Object}
can safely read an \code{Array},
but not the reverse.
We call~\textit{host code}
the code that the host runs
and checks, such as the script.
Host code is~\emph{type-safe}~\cite{Cardelli97}:
the host checks the tag of every operand
and raises an exception if it is wrong.

\point{Native code and binding API}
The script imports \code{ext},
a \emph{native extension},
i.e., a compiled module,
such as a Node.js addon or a CPython extension module,
typically written in C, C++ or Rust.
We call its code~\emph{native code}.
The extension registers \emph{native entry points} with the host,
i.e., native functions that scripts can call,
such as \code{Total} (lines 1--10),
which the script calls as \code{ext.total}.
Native code consists of the entry points
and the auxiliary native functions they call.
It shares the address space of the host
and accesses the memory of host values directly,
without any checks,
so native code is not type-safe.
Native code operates on host values
through the \emph{binding API} of the host,
such as CPython's C-API~\cite{python-capi}
and Node.js's N-API~\cite{node-api},
or libraries that wrap them,
such as {\tt node-addon-api}~\cite{node-addon-api}
and {\tt pybind11}~\cite{pybind11}.
We call a host value
that native code has access to
a \emph{boundary value}.
The arguments of an entry point are boundary values,
such as \code{a} and \code{f}, which line~2 reads.
Native code receives each of them
as a handle of a general type,
such as \code{Value} in V8 or \code{PyObject *} in CPython,
which reveals nothing about its tag.

\point{Tests, uses and host operations}
Through the binding API,
native code can test or use a boundary value.
A \emph{test} inspects the tag of the value,
e.g., line~3 tests whether \code{a} is an \code{Array}.
A \emph{use} accesses the memory of the value
according to one fixed layout,
and thus assumes that the value has
the tag that determines that layout,
e.g., line~4 reads the length of \code{a},
so it assumes that \code{a} is an \code{Array}.
A test \emph{protects} a use
if the use runs only when the test succeeds,
and every value that passes the test
has the tag that the use assumes
or a tag below it.
For example,
the test on line~3 protects the use on line~4.
Native code can also ask the host
to perform an operation on a value,
such as a member read, a call or a conversion.
We call such a request a \emph{host operation}.
Its result is a~\emph{new} boundary value.
Since scripts can customize these operations,
e.g., through getters or conversion methods
such as \code{toString},
this result can be any value
(``anything'' in the figure).
For example,
line~6 calls the function \code{() => "x"}
that the script passed to \code{total},
which returns a string
but could just as well return a number.
A host operation can also fail.
When host code raises an exception,
native code receives an empty value,
which line~7 tests for.

\point{Most-general client}
Packages may validate values in script code
before calling native entry points,
but such validation is often missing~\cite{flowjs}
or can be bypassed~\cite{brown-checker},
and any script can import an extension.
For the rest of the paper,
we assume
that an entry point can receive any value~\cite{KM19}
(\textit{anything} in the figure),
and that a call back into the host
can run arbitrary script code.

\point{Boundary type errors}
A~\emph{boundary type error} occurs
when native code accesses a value
assuming a tag (and thus a layout)
that the value does not have.
For example,
line~8 reads~\code{b} as a number,
although no test establishes that it is one,
so \code{total} returns garbage to the script.

\subsection{Motivating Examples}
\label{sec:motivating}
We discuss two boundary type errors
that our approach uncovered in widely used native extensions
(Figure~\ref{fig:motivating}).
We use them to derive our key insights
in \cref{sec:insights}.

\point{Segmentation fault in Node.js canvas}
The {\tt canvas} package~\cite{canvas} provides
2D drawing for Node.js.
It is written against {\tt node-addon-api}~\cite{node-addon-api}
and has more than 40 million monthly downloads.
Figure~\ref{fig:motivating-a} shows
how two property assignments cause a process crash.
The script assigns
a plain object to \code{cv.context} (line~S3),
then sets \code{cv.width = 20} (line~S4).
This invokes the native setter \code{Canvas::SetWidth}
(lines 11--14),
which calls the auxiliary \code{resurface}
(line~13).
\code{resurface} reads the property \code{context}
through the host operation \code{Get} (line~2),
which may run a getter that the script defined.
Then,
\code{resurface} checks
that the read did not fail
(\code{IsEmpty})
and that the result~\code{ctx} is an object
(\code{IsObject}, line~3).
On line~5,
\code{resurface} then treats~\code{ctx}
as a \code{Context2d},
a class that the extension defines,
and reads the native data stored in its layout
(\code{Unwrap}).

The problem is that the check on line~3 is too permissive.
It accepts any object,
including \code{\{\}}
that the script assigned to \code{cv.context}
(line~S3).
However,
the layout of a plain object is not compatible
with that of a \code{Context2d},
so \code{Unwrap} obtains an invalid pointer,
and the process crashes
with a segmentation fault.
The fix
(lines~4 and~8)
checks that~\code{ctx} is
a \code{Context2d} before using it.

\point{Silent output truncation in Python's built-in \code{io}}
Figure~\ref{fig:motivating-b} shows
a boundary type error in \code{\_io},
a native extension of CPython's standard library,
that silently truncates
text written to a file.
Its entry point \code{write} (lines~1--11)
calls the file's encoder through a host operation (line~2)
and expects its result~\code{b} to be a byte sequence (line~4).
As an optimization for ASCII text,
which needs no encoding,
\code{write} also accepts the text object itself
(\code{b == text}, line~4)
and then copies its characters to the output,
one byte per character (lines~8--10),
which is correct only for ASCII text.

However,
nothing enforces this assumption.
A custom encoder can return
the given text unchanged
even when it is not ASCII
(and thus \code{b == text}).
The script in Figure~\ref{fig:motivating-b}
registers such an encoder
(lines~S2--S3),
opens a file with it (line~S4),
and writes \code{"DELETE: NO\texttrademark"} 
(line~S5).
Because of the non-ASCII character \texttrademark,
CPython stores this string
with two bytes per character,
yet line~10 copies one byte per character,
so reading the file back (line~S7) yields only \code{DELETE}.
The fix (lines~6--7)
checks that~\code{b} is ASCII
whenever it is the original text object.

\subsection{Key Insights and State of the Art}
\label{sec:insights}

The examples above lead to four
observations that underpin our design.

\point{I1: Tags determine type safety in native code}
Whether a use is type-safe depends
only on the tag of the boundary value it accesses,
since the tag determines the value's memory layout.
For example,
in Figure~\ref{fig:motivating-a},
the layout of the object \code{\{\}} is incompatible with
that of a \code{Context2d}.
Therefore,
detecting such errors requires
\emph{reasoning about the tags of boundary values}.

\point{I2: Tags form a hierarchy}
Tags are ordered:
\code{Array} lies below \code{Object},
since every array is also an object.
A test protects a use only
if every tag it accepts
lies below the tag the use assumes.
In Figure~\ref{fig:motivating-a},
line~3 tests
for an object,
but line~5 assumes the narrower \code{Context2d},
so a plain object passes the test,
leading to a segmentation fault.
Detecting such errors requires
\emph{knowing the tag hierarchy},
including the tags of the classes
that extensions register.
For example,
{\tt canvas} registers \code{Context2d} with the host
when \code{require(\textquotesingle canvas\textquotesingle)}
loads its native extension.

\point{I3: Only tests establish tags}
Native code can encode
implicit beliefs about
a boundary value's tag in branches on native data.
For example,
in Figure~\ref{fig:motivating}b,
line~4 compares~\code{b} with~\code{text},
and the developer assumes that~\code{b} is ASCII
whenever the two are the same object.
However,
this comparison does not examine the tag of~\code{b},
so nothing guarantees this assumption at run time.
An approach for detecting such type errors
must \emph{learn tags only from explicit tag tests}.

\point{I4: Only tags are stable}
A host operation runs script code,
which is arbitrary.
Its result can be any value,
even if the same host operation returned
a well-typed value before.
The script code can also modify
the contents of the values that native code holds,
i.e., their raw bits in Figure~\ref{fig:interaction},
e.g., by replacing the object members or
array elements.
However,
it cannot change their tags,
and thus their layouts.
For example,
a script can define a~\emph{getter} for a property \code{o.f},
i.e., a script function
that runs every time the property is read
(\cref{app:examples} discusses
a concrete example).
Such a getter may return
an array the first time
and a number the second time,
while~\code{o} itself remains the same object.
Therefore,
\emph{every value that native code obtains from the host
can be any value},
regardless of what earlier operations returned.


\point{Limitations of existing work}
Existing approaches cannot detect the
errors of Figure~\ref{fig:motivating}.
The checker of
Brown et al.~\cite{brown-checker}
and FlowJS~\cite{flowjs}
track the arguments of a native
function to their uses,
and consider a use safe as soon as
any test of the argument lies on the path.
Since they do not reason about tags
(\textbf{I1} and \textbf{I2}),
both would accept the test on line~3
of Figure~\ref{fig:motivating-a}
as a guard for the use on line~5.
Furthermore,
neither treats the read of
\code{context} on line~2
as a new value that needs its own test
(\textbf{I4}).
Furr and Foster~\cite{FF05,FF08} target OCaml and Java,
where the host program declares
the type of each value it passes to native code,
e.g., a Java method \code{native int f(String s)}
guarantees that~\code{s} is a string.
They infer the type
that native code assumes for each value
and check it against this declaration.
JavaScript and Python declare no such types,
so only the tests of native code
can establish a tag (\textbf{I3}).
MOPSA~\cite{mopsa}
analyzes a Python program
with its C extensions
for runtime errors,
such as invalid memory accesses,
but does not reason about tags
(\textbf{I1} and \textbf{I2}).

\point{Tag safety}
The host of dynamic languages
provides no guarantees about
the values it passes to native code.
Our central idea is
that the type safety of native code
then reduces to \emph{tag safety}:
\textit{every use is protected by a test
that guarantees the tag the use assumes}.
Unlike prior work on statically typed languages,
checking it requires neither declared types
nor knowledge of the script.
One might expect to also need
structural and higher-order types,
i.e.,
the types of object members
and function results,
e.g.,
that member~\code{f} of an object is a string.
Inferring such types for dynamically
typed languages is hard.
Our insight is that we do not need them:
every member read and every function call
is itself a host operation,
whose result is a new boundary value
that native code must test on its
own before using it
(\textbf{I3} and \textbf{I4}).

\section{Problem Formulation}
\label{sec:approach}

To effectively reason about
boundary type errors,
we introduce a simple language
that models native code,
and formalize them over it.

\point{The XLIR language}
The insights of \cref{sec:insights}
motivate the design of XLIR,
a small language that abstracts
each boundary value to its tag.
The values of XLIR are tags,
together with a special value
$\emp$ for a failed host operation
(\textbf{I1}).
Each program carries the host's tag hierarchy,
so a use that assumes a tag
also accepts every tag below it (\textbf{I2}).
XLIR keeps the tag tests of native
code and models every other branch as a
nondeterministic choice~$\star$ (\textbf{I3}).
Finally,
XLIR does not model the members
of host values or writes to them.
Since member reads can invoke script-defined code,
each read is a host operation
that returns an arbitrary value (\textbf{I4}).

An XLIR program $P=(e,\bar F,\mathcal{T})$
(Figure~\ref{fig:xlir}) consists of
the name $e$ of a native entry point,
a set $\bar F$ of native function
definitions that also includes the
definition of $e$,
and a tag hierarchy $\mathcal{T}$.
We call $P$ the \emph{program of} $e$.
The entry point has no parameters:
it reads the arguments
that the host passes to it
through dedicated syntax,
i.e.,
$\Let{x}{\Arg{i}}$.
Every other function receives
values through its parameters.
Calls and returns only pass
values between functions,
so every value of a run comes
from an argument of the entry
point or from a host operation.
XLIR is in SSA form\label{def:variable},
so within a function call a variable
is bound once, either as a parameter or by a $\kw{let}$.

\begin{figure}[t]
\small
\newcommand{\mv}[2]{\langle #1\in\mathrm{#2}\rangle}
\captionsetup[sub]{justification=centering,singlelinecheck=false}
\newcommand{\setof}[1]{\text{#1}}
\newsavebox{\xlirsyn}\savebox{\xlirsyn}{\begin{minipage}[t]{0.55\linewidth}
\vspace{0pt}
\centering
{\renewcommand{\arraystretch}{1.12}%
$\begin{array}{@{}l@{\;\;}c@{\;\;}l@{}}
\mv{P}{Program} & ::= & (e,\bar F,\mathcal{T}) \\
\mv{F}{Func} & ::= & \kw{fun}\ f(\bar x)=\sq \\
\mv{s}{Stmt} & ::= & \Let{x}{\Arg{i}}\\
  & \mid & \Let{y}{\Host{\tau}{x}}\\
  & \mid & \proj{\tau}{x}\\
  & \mid & \kw{if}\ \ischk{\tau}{x}\ \kw{then}\ \sq_1\ \kw{else}\ \sq_2\\
  & \mid & \kw{if}\ \star\ \kw{then}\ \sq_1\ \kw{else}\ \sq_2\\
  & \mid & \Let{\bar y}{f(\bar x)}\\
  & \mid & \kw{return}\ \bar x\\
\mv{\sq}{Stmt^*} & ::= & \eseq \mid s;\sq\\
\mv{x,y}{Var} & ::= & \setof{variable names}\\
\mv{f,e}{FName} & ::= & \setof{function names}\\
\mv{\mathcal{T}}{Hier} & ::= & \setof{tag hierarchies}
\end{array}$}
\subcaption{Syntax of XLIR.}\label{fig:syntax}
\end{minipage}}%
\newlength{\xlirh}\setlength{\xlirh}{\dimexpr\ht\xlirsyn+\dp\xlirsyn\relax}%
\usebox{\xlirsyn}\hfill
\begin{minipage}[t][\xlirh][t]{0.43\linewidth}
\vspace{0pt}
\centering
\vspace*{\fill}
{\renewcommand{\arraystretch}{1.0}%
$\begin{array}{@{}l@{\;\;}c@{\;\;}l@{}}
\mv{v}{Val} & ::= & t \mid \emp\\
\mv{t}{\Hs{Tag}} & ::= & \tnum \mid \tstr \mid \AsciiStr\\
  & \mid & \kw{Array} \mid \kw{Object} \mid \dots
\end{array}$}
\subcaption{Values of XLIR.}\label{fig:values}
\vfill
{\renewcommand{\arraystretch}{1.0}%
$\begin{array}{@{}l@{\;\;}c@{\;\;}l@{}}
\mv{\sigma,\tau}{Type} & ::= & \Any \mid \None \mid t \mid \exact{t}\\
  & \mid & \tau\tor\tau \mid \tau\wedge\tau \mid \neg\tau\\[3pt]
\sem{t} & = & \{t'\in\Hs{Tag}\mid t'\preceq t\}\\
\sem{\exact{t}} & = & \{t\}\\
\sem{\None} & = & \{\emp\}\\
\sem{\Any} & = & \mathrm{Val}\\
\sem{\tau_1\tor\tau_2} & = & \sem{\tau_1}\cup\sem{\tau_2}\\
\sem{\tau_1\wedge\tau_2} & = & \sem{\tau_1}\cap\sem{\tau_2}\\
\sem{\neg\tau} & = & \mathrm{Val}\setminus\sem{\tau}
\end{array}$}
\subcaption{Types in XLIR and their denotations.}\label{fig:types}
\end{minipage}
\figrule
\vspace{-8pt}
\caption{The syntax, values and types of XLIR.
The sets marked $H$ belong to the host.}
\label{fig:xlir}

\vspace{-6pt}
\end{figure}

\begin{wrapfigure}{r}{0.47\linewidth}
\vspace{-0.8\baselineskip}
\centering
\resizebox{\linewidth}{!}{\begin{tikzpicture}[x=1.0cm, y=1.0cm, every node/.style={inner sep=1.2pt, font=\small},
  e/.style={draw=black!55, line width=0.4pt}]
\node (obj) at (0,1)    {$\kw{Object}$};
\node (num) at (1.25,1) {$\tnum$};
\node (str) at (2.25,1) {$\tstr$};
\node (bool) at (3.25,1) {$\tbool$};
\node (und) at (4.3,1)  {$\tundef$};
\node (rd)  at (5.4,1) {$\dots$};
\node (arr) at (-1.0,0) {$\kw{Array}$};
\node (fun) at (0.3,0)  {$\kw{Function}$};
\node (c2d) at (1.85,0) {\textcolor{regdraw}{$\kw{Context2d}$}};
\node (od)  at (3.05,0) {$\dots$};
\foreach \n in {arr,fun,c2d,od} \draw[e] (obj) -- (\n);
\end{tikzpicture}}
\caption{The tag hierarchy $\mathcal{T}$ of the program of
\cref{fig:motivating-a}.
The \code{canvas} extension registers
\textcolor{regdraw}{$\kw{Context2d}$},
and the other tags are built into the host (V8).}
\label{fig:tag-hierarchy}
\vspace{-0.8\baselineskip}
\end{wrapfigure}

\point{Statements (Figure~\ref{fig:syntax})}
$\Let{x}{\Arg{i}}$ reads argument
$i$ of the entry point.
$\Let{y}{\Host{\tau}{x}}$ is a host
operation on $x$ that returns an arbitrary
value of type $\tau$.
$\Ifs{\ischk{\tau}{x}}{\sq_1}{\sq_2}$ is a tag test.
$\proj{\tau}{x}$ is a use,
which assumes that $x$ has type $\tau$.
$\Ifs{\star}{\sq_1}{\sq_2}$ is a branch on native data.
Finally,
$\Let{\bar y}{f(\bar x)}$ calls
the native function $f$,
and we write
$f(\bar x)$ when $\bar y$ is empty.

\point{Values (Figure~\ref{fig:values})}
\label{def:value}\label{def:tag}%
A value of XLIR is either a tag or
the special value $\emp$,
which a host operation returns
when it fails
(e.g., it throws an exception).
The tag hierarchy $\mathcal{T}$ of
a program is a set of tags with a
partial order $\preceq$.
It contains the host's built-in tags
and the tags of the classes
that the native
extension registers through the binding API,
such as $\kw{Context2d}$ in Figure~\ref{fig:motivating-a}
(\cref{fig:tag-hierarchy}).
A script can create tags only below existing ones
(e.g., a Python subclass of \code{list}),
so the tag of every value lies below a
tag in $\mathcal{T}$.

\point{Types (Figure~\ref{fig:types})}
\label{sec:approach-sub}\label{def:type}\label{def:testable}\label{def:den}\label{lem:sub}%
Whereas a tag belongs to a value at run time,
a type~$\tau$ is static:
it denotes a set of values~$\sem{\tau}$,
following the
set-theoretic approach to types~\cite{FCB08}.
Types label host operations, tests and uses,
and tag refinement (\cref{sec:tag-refinement})
assigns one to each variable at each program point.
A value $v$ has type $\tau$,
written $\vdash v:\tau$,
when $v\in\sem{\tau}$.
The type
$\Any$ denotes all values,
and $\None$ denotes only the special value $\emp$.
A tag $t$,
as a type,
denotes $t$ and every tag below it.
For example,
$\sem{\kw{Object}}=\{\kw{Object},\kw{Array},\kw{Function},\kw{Context2d},\dots\}$.
The exact type $\exact{t}$ denotes only $t$;
it models tests such as
\code{Py\_IS\_TYPE}, which reject
the tags below~$t$.
A union $\tau_1\tor\tau_2$ denotes
the values of either type, an
intersection $\tau_1\wedge\tau_2$ the values of both,
and a negation $\neg\tau$
the values outside $\sem{\tau}$.

\newlength{\savedintextsep}\setlength{\savedintextsep}{\intextsep}\setlength{\intextsep}{4pt}%
\begin{wrapfigure}{r}{0pt}
\vspace{-0.5\baselineskip}
\footnotesize
$\begin{array}[t]{@{}r@{\;\;}l@{}}
&\kw{fun}\ \mathit{write}()=\\
\xln{1}&\quad \Let{x}{\Arg{1}};\\
\xln{2}&\quad \Let{b}{\Host{\Any}{x}};\\
\xln{3}&\quad \kw{if}\ \ischk{\None}{b}\ \kw{then}\ \kw{return}\ \kw{else}\ \eseq;\\
\xln{4}&\quad \kw{if}\ \star\ \kw{then}\ (\kw{if}\ \ischk{\tbytes}{b}\ \kw{then}\ \eseq\ \kw{else}\ \kw{return})\\
\xln{5}&\quad \kw{else}\ \fixbox{(\kw{if}\ \ischk{\AsciiStr}{b}\ \kw{then}\ \eseq\ \kw{else}\ \kw{return})};\\
\xln{6}&\quad \kw{if}\ \ischk{\tstr}{b}\ \kw{then}\\
\xln{7}&\quad \quad\proj{\tstr}{b};\\
\xln{8}&\quad \quad\proj{\AsciiStr}{b}\\
\xln{9}&\quad \kw{else}\ \eseq;\\
\xln{10}&\quad \kw{return}
\end{array}$
\caption{XLIR program of \code{write} (\cref{fig:motivating-b}).}
\label{fig:xlir-ex}\label{fig:xlir-ex-b}
\vspace{-0.6\baselineskip}
\end{wrapfigure}
\point{Example}
Figure~\ref{fig:xlir-ex} shows the
program of \code{write}
(Figure~\ref{fig:motivating-b}) in XLIR.
Section~\ref{sec:lowering} explains
how binding API calls map to XLIR statements.
Line~2 is a host operation
that calls the encoder,
so $b$ may be any value,
and line~3 tests whether it is $\emp$.
Line~4 is the pointer comparison
\code{b != text},
which XLIR writes as $\star$.
Line~6 tests that $b$ is a $\tstr$,
and lines~7 and~8 are uses that
assume $\tstr$ and $\AsciiStr$.
Line~5 is the fix,
a test for $\AsciiStr$.

\setlength{\intextsep}{\savedintextsep}%
\point{Semantics}
We give XLIR a small-step semantics.
A \emph{configuration} $\Cfg{\env}{\sq}\cdot K$
is a stack of frames,
one per active function call,
with $\Cfg{\env}{\sq}$ on top.
A frame holds an environment~$\env$,
which maps variables to values,
and the statements~$\sq$ that remain to run.
A run of~$P$ starts at the entry point
with a vector of argument tags,
and ends when the entry point returns.
Host operations and choices~$\star$
are nondeterministic:
a host operation may return any value of its type,
and a choice may take either branch.
A use stops the run in the state~$\wrong$
when the value it accesses
does not have the type it assumes:
\[
\inferrule*[lab=S-UseUnsafe]{\not\vdash\env(x):\tau}
 {\Cfg{\env}{\proj{\tau}{x};\sq}\cdot K\to\wrong}
\]
The remaining rules are standard,
and \cref{app:semantics} gives them in full.

\begin{defn}[Boundary type error]
\label{def:bte}
A program point with $\proj{\tau}{x}$
in $P$ is a \emph{boundary type
error} if a run of $P$ reaches
it in a configuration
$\Cfg{\env}{\proj{\tau}{x};\sq}\cdot K$
with $\not\vdash\env(x):\tau$,
that is,
$\env(x)\notin\sem{\tau}$.
A program is \emph{tag-safe} if it has none.
\end{defn}


\section{Practical Identification of Boundary Type Errors}
\label{sec:practical}

\begin{figure}[t]
\centering
\includegraphics[scale=0.37]{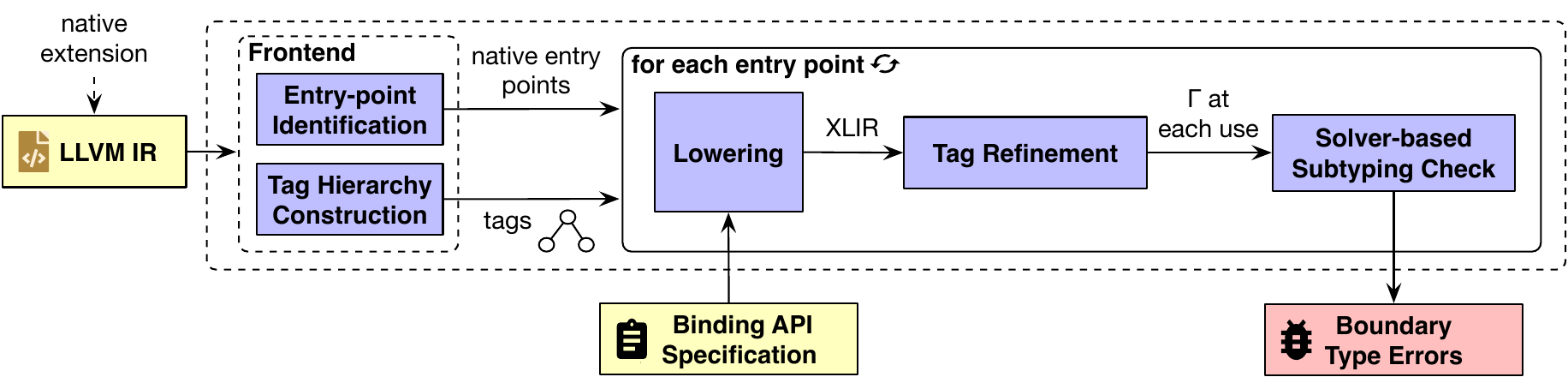}
\vspace{-8pt}
\caption{The high-level overview of our approach
for detecting boundary type errors in native extensions.}
\label{fig:arch}
\vspace{-6pt}
\end{figure}

\point{Overview}
Figure~\ref{fig:arch} presents
a high-level overview of our approach.
Since native extensions are written in C,
C++ or Rust,
the approach analyzes their common compilation target,
LLVM IR,
in four phases.
First,
the \emph{frontend}
identifies the extension's entry points
and builds its tag hierarchy.
For each entry point,
\emph{lowering} slices the program
by following its boundary values
to every LLVM instruction they reach,
and translates these instructions into an XLIR program,
using a binding API specification
that states what each binding API function does.
Then,
\emph{tag refinement},
a flow-sensitive dataflow analysis,
computes a type for each XLIR variable
at each point of the program.
Finally,
at every use $\proj{\tau}{x}$,
an SMT solver checks whether the type
that tag refinement computed for~$x$
is a subtype of~$\tau$,
and if not,
reports the use with a witness value
(\cref{sec:smt}).

\subsection{Frontend}
\label{sec:entry-point}

\begin{table}[t]
\caption{Representative calling conventions of entry points, one rule per row.
The grey label under each kind is its registration point.
Bold parameters carry boundary values, and
$\bar{x}$ stands for $x_0, x_1, \dots$.}
\label{tab:entries}
\vspace{-5pt}
\scriptsize
\renewcommand{\arraystretch}{0.9}%
\setlength{\tabcolsep}{3pt}
\renewcommand{\code}[1]{{\scriptsize\texttt{#1}}}%
\newcommand{\sig}[2]{{\scriptsize\ttfamily\begin{tabular}[t]{@{}>{\scriptsize\ttfamily}l@{}}#1(\\\hspace*{1em}\begin{tabular}[t]{@{}>{\scriptsize\ttfamily}l@{}}#2)\end{tabular}\end{tabular}}}%
\newcommand{\siga}[2]{{\scriptsize\ttfamily#1(\begin{tabular}[t]{@{}>{\scriptsize\ttfamily}l@{}}#2)\end{tabular}}}%
\newcommand{\bv}[1]{\textbf{#1}}%
\newcommand{\apitag}[1]{\tikz[baseline=(t.base)]\node[fill=black!8,rounded corners=1pt,inner xsep=1.5pt,inner ysep=1pt](t){\scriptsize\ttfamily\strut#1};}%
\newcommand{\kind}[2]{\begin{tabular}[t]{@{}l@{}}#1\\\apitag{#2}\end{tabular}}%
\newcommand{\lines}[1]{\begin{tabular}{@{}l@{}}#1\end{tabular}}%
\newcommand{\softrule}{\arrayrulecolor{black!25}\midrule\arrayrulecolor{black}}%
\begin{tabularx}{\linewidth}{@{}llll>{\raggedright\arraybackslash}X@{}}
\toprule
\textbf{Binding API} & \textbf{Kind} & \textbf{Host call} & \textbf{C/C++ signature} & \textbf{Boundary values}\\
\midrule
NAN & \kind{method}{SetPrototypeMethod} & $o.f(\bar{x})$
  & \sig{void f}{const Nan::FunctionCallbackInfo\\<Value>\& \bv{info}}
  & \mbox{$o$ = \code{\bv{info}.This()}}, \mbox{$x_k$ = \code{\bv{info}[k]}}\\
\softrule
\multirow{5}{*}{\lines{node-\\addon-api}}
  & \kind{getter}{InstanceGetter\ldots} & $o.p$
  & \sig{Napi::Value T::g}{const Napi::CallbackInfo\& \bv{info}}
  & \mbox{$o$ = \code{\bv{info}.This()}}\\
  & \kind{setter}{InstanceSetter\ldots} & $o.p = v$
  & \sig{void T::s}{const Napi::CallbackInfo\& \bv{info},\\const Napi::Value\& \bv{v}}
  & \mbox{$o$ = \code{\bv{info}.This()}}, \mbox{$v$ = \code{\bv{v}}}\\
\softrule
\multirow{6}{*}{CPython}
  & \kind{method}{METH\_O} & $o.f(x)$
  & \multirow{6}{*}{\siga{PyObject *f}{PyObject *\bv{p0},\\PyObject *\bv{p1}}}
  & \mbox{$o$ = \code{\bv{p0}}}, \mbox{$x$ = \code{\bv{p1}}}\\
  & \kind{variadic}{METH\_VARARGS} & $o.f(\bar{x})$ & & \mbox{$o$ = \code{\bv{p0}}}, \mbox{$x_k$ = \code{PyTuple\_GET\_ITEM(\bv{p1},k)}}\\
  & \kind{operator}{nb\_add} & $a + b$ & & \mbox{$a$ = \code{\bv{p0}}}, \mbox{$b$ = \code{\bv{p1}}}\\
\bottomrule
\end{tabularx}
\vspace{-6pt}
\end{table}

The frontend relies on one observation:
an extension must register its entry points
and native classes with the host.
Entry-point identification finds where
the extension registers each entry point,
which also determines how the host passes its arguments.
Tag hierarchy construction finds where
it registers its native classes,
and places them among the host's built-in tags.

\point{Signatures}
To find entry points,
our approach exploits the fact that
each binding API dictates their signature
(Table~\ref{tab:entries}).
For example,
every entry point written against NAN
takes a single \code{Nan::FunctionCallbackInfo<Value>\&}
that holds the call's receiver and arguments,
and returns \code{void}.
Therefore,
our approach scans the given LLVM module
and identifies as entry points
all native functions whose signature
matches one that the binding API
of the extension specifies.

\point{Challenge}
A signature alone does not determine
how the host passes the arguments.
For example,
CPython uses native functions of signature
\code{PyObject *f(PyObject *p0, PyObject *p1)}
in several ways (Table~\ref{tab:entries}).
For a method that takes one argument
(\code{METH\_O}),
when a script evaluates \code{o.f(x)},
CPython calls~\code{f}
with~\code{o} as~\code{p0}
and~\code{x} as~\code{p1}.
For a method that takes any number of arguments
(\code{METH\_VARARGS}),
\code{p1} is a tuple holding all of them.
For the operator~\code{+} (\code{nb\_add}),
\code{p0} and \code{p1} are its two operands.
Confusing these conventions causes
both false positives and false negatives.
Consider \code{PyTuple\_GET\_ITEM(p1, k)},
which reads the $k$-th element of~\code{p1}
assuming that~\code{p1} is a tuple.
In a \code{METH\_VARARGS} function,
this read is safe,
since CPython always passes a tuple as~\code{p1}.
In a \code{METH\_O} function,
\code{p1} is whatever value the script passes,
so the read is a boundary type error.

\point{Registration points}
While the signature of an entry point does not determine
how the host passes its arguments,
its registration does.
To expose a native function to scripts,
the extension registers it with the host,
stating under which name scripts call it
and how the host should pass the arguments.
A registration takes one of two forms.
(1)~The extension calls
a registration function of the binding API.
For example,
\code{canvas} calls \code{InstanceAccessor}
of {\tt node-addon-api}
to register \code{SetWidth} (\cref{fig:motivating-a})
as the setter of the property \code{width}.
(2)~The extension stores the entry point
in a table that the host reads.
For example,
\code{\_io} lists the methods of its file objects
in a table with one entry per method,
such as \code{\{"write", write, METH\_O\}}.
The string \code{"write"} is the name
under which scripts call the method,
\code{write} is the native function that implements it
(\cref{fig:motivating-b}),
and the flag \code{METH\_O} states
that CPython passes the single argument of the method
as the second parameter of~\code{write}.
Registrations typically take place
in the module's initialization function
or in global tables
(see ``Limitations'' in Section~\ref{sec:implementation}),
so our approach scans the entire module for them
and extracts the calling convention of
each entry point from its registration.

\point{Entry-point identification}
Our approach identifies entry points
with rules
for each binding API it supports.
Each rule pairs a signature
with a kind of registration point
(e.g., \code{METH\_O})
and states how the host passes the arguments
to such an entry point.
Each row of Table~\ref{tab:entries} is a rule.
\Cref{app:conventions} lists all the rules.

\point{Tag hierarchy construction}
The host defines a fixed set of built-in tags
for its own values,
such as \code{Number}, \code{String}, \code{Array} and \code{Object} in V8,
together with their order.
We encode these tags once per host.
We also treat a distinction that the host records
in a value's header as a tag,
such as CPython's flag for ASCII strings,
which selects the layout of the string's characters.
An extension can add its own tags
by registering native classes.
While scanning for registrations,
our approach also collects these classes,
such as \code{Context2d},
which \code{canvas} registers
with \code{ObjectWrap<Context2d>::DefineClass},
and adds each as a new tag.
The built-in tags and the extension's tags
together form the tag hierarchy~$\mathcal{T}$.

\subsection{Lowering to XLIR}
\label{sec:lowering}

Lowering turns
each entry point,
together with the native functions it calls,
into an XLIR program.
Starting from the entry point's boundary values
(Section~\ref{sec:entry-point}),
it computes a~\emph{forward slice} of the module
that keeps only the instructions relevant to type safety:
(1) those that obtain boundary values,
by reading an argument of the entry point
or through a host operation
(Section~\ref{sec:prelims}),
(2) those that test or use such a value,
and (3) the branches that decide which of them execute.
All other instructions,
such as native arithmetic,
cannot cause a boundary type error and are ignored.
Lowering computes the slice
on the module's sparse value-flow graph (SVFG).
Unlike LLVM's def-use chains,
the SVFG follows values through memory
and across function calls,
including indirect ones.
This means that following the edges of this graph
from the boundary values
reaches every instruction these values may reach.
However,
the slice does not determine in which order
these instructions run,
or which branches guard them.
Lowering takes this information
from the module's control-flow graph (CFG).

\begin{table}[t]
\caption{The binding API specification~$\mathcal{K}$
for the binding API functions of \cref{fig:motivating}.
$T$ is the class that \code{ObjectWrap<T>} wraps.
Position~0 of a method is its receiver.}
\label{tab:spec}
\vspace{-7pt}
\centering
\footnotesize
\renewcommand{\code}[1]{{\scriptsize\texttt{#1}}}%
\begin{tabular}{@{}ll@{\hspace{2em}}ll@{}}
\toprule
\textbf{node-addon-api}~$\boldsymbol{f}$ & $\boldsymbol{\mathcal{K}(f)}$
  & \textbf{CPython}~$\boldsymbol{f}$ & $\boldsymbol{\mathcal{K}(f)}$\\
\midrule
\code{Object::Get} & $\{\langle\kw{hop},\Any,\code{ret}\rangle,$
  & \code{PyObject\_CallOneArg} & $\{\langle\kw{hop},\Any,\code{ret}\rangle\}$\\
 & $\phantom{\{}\langle\kw{use},\neg\tnull\wedge\neg\tundef,0\rangle\}$
  & \code{PyBytes\_Check} & $\{\langle\kw{is},\tbytes,0\rangle\}$\\
\code{Value::IsEmpty} & $\{\langle\kw{is},\None,0\rangle\}$
  & \code{PyUnicode\_Check} & $\{\langle\kw{is},\tstr,0\rangle\}$\\
\code{Value::IsObject} & $\{\langle\kw{is},\AnyObj,0\rangle\}$
  & \code{PyUnicode\_IS\_ASCII} & $\{\langle\kw{is},\AsciiStr,0\rangle\}$\\
\code{ObjectWrap<T>::Unwrap} & $\{\langle\kw{use},T,0\rangle\}$
  & \code{PyUnicode\_GET\_LENGTH} & $\{\langle\kw{use},\tstr,0\rangle\}$\\
 &
  & \code{PyUnicode\_1BYTE\_DATA} & $\{\langle\kw{use},\AsciiStr,0\rangle\}$\\
\bottomrule
\end{tabular}
\vspace{-6pt}
\end{table}

\point{Binding API specification}
\label{def:spec}
In LLVM IR,
a call to the binding API is a call to an external function,
so the IR does not show
what the call does with boundary values,
i.e., whether it tests their type,
assumes it,
or runs host code.
Lowering takes this knowledge
from a~\emph{binding API specification}~$\mathcal{K}$,
written once per binding API.
Formally,
\[
\mathcal{K} : \mathit{FName} \to \mathcal{P}(\mathit{Kind} \times \mathit{Type} \times \mathit{Pos}),
\]
where $\mathit{Kind} = \{\kw{is}, \kw{use}, \kw{hop}\}$
and $\mathit{Pos} = \mathbb{N} \cup \{\code{ret}\}$.
$\mathcal{K}$ maps
each binding API function~$f$
to the set of effects that a call to~$f$ has.
For example,
an effect
$\langle \kw{is}, \tau, j \rangle \in \mathcal{K}(f)$
states that~$f$ tests
whether its argument~$j$ has type~$\tau$,
$\langle \kw{use}, \tau, j \rangle$
states that~$f$ assumes it,
and $\langle \kw{hop}, \tau, \code{ret} \rangle$
states that~$f$ performs a host operation
and returns a value of type~$\tau$.
Table~\ref{tab:spec} gives~$\mathcal{K}$
for the binding API functions
that appear in Figure~\ref{fig:motivating}.
A function can have several effects.
For example,
\code{Object::Get} both assumes that its receiver
is neither \code{undefined} nor \code{null},
as required by ECMAScript property access,
and reads a property of it through a host operation,
whose result can be any value.

Given a binding API specification~$\mathcal{K}$
and the module's SVFG and CFG,
lowering produces the XLIR program
in a single pass over the SVFG.
Figure~\ref{fig:lowering-canvas} shows
the simplified LLVM IR of Figure~\ref{fig:motivating-a}
(without the fix),
its SVFG,
and the XLIR program of \code{SetWidth}.

\newsavebox{\lcListing}\newsavebox{\lcGraph}\newsavebox{\lcXlir}
\newlength{\lcHeight}
\begin{figure}[t]
\centering
\renewcommand{\theFancyVerbLine}{\textcolor{black!40}{\sffamily\fontsize{6}{7}\selectfont\arabic{FancyVerbLine}}}%
\setminted{fontsize=\fontsize{6.6}{7.5}\selectfont, xleftmargin=12pt, numbersep=5pt, escapeinside=||, formatcom=\codepalette}%
\captionsetup[sub]{justification=centering,singlelinecheck=false,font=footnotesize,position=below,skip=2pt}%
\colorlet{entryfill}{black!9}%
\tikzset{
  vf/.style={draw=black!60, rounded corners=1.5pt, font=\ttfamily\scriptsize,
    inner sep=1.6pt, minimum height=3.8mm, fill=white},
  vfstart/.style={vf},
  vfarg/.style={vf, fill=argfill},
  vftest/.style={vf, fill=weakfill},
  vfuse/.style={vf, fill=badfill},
  vfreg/.style={-{Stealth[length=1.4mm]}, line width=0.4pt},
  vfmem/.style={vfreg, dash pattern=on 2pt off 1.2pt},
  vfcall/.style={vfreg, dotted, line width=0.6pt},
  vflab/.style={font=\ttfamily\tiny, fill=white, inner sep=0.5pt},
  vforder/.style={circle, fill=black!70, text=white, inner sep=0pt,
    minimum size=2.5mm, font=\sffamily\bfseries\fontsize{5}{5}\selectfont},
}%
\newcommand{\vforder}[2]{\node[vforder] at (#1.north west) {#2};}%
\newcommand{\vfo}[1]{\tikz[baseline=(o.base)]\node[vforder](o){#1};}%
\newcommand{\lcLeft}{0.39\textwidth}%
\newcommand{\lcMid}{0.28\textwidth}%
\newcommand{\lcRight}{0.33\textwidth}%
\begin{lrbox}{\lcListing}\begin{minipage}[t]{\lcLeft}
\begin{minted}[linenos]{llvm}
|\hb{entryfill}|define @SetWidth(%info) {|\he|
  |\hb{argfill}|%t = call |\api{@This}|(%info)|\he|
  %o = call |\api{@As\textless Object\textgreater}|(%t)
  call @resurface(%o)
}
define @resurface(%This) {
  ...
  store %This, %obj
  call |\api{@Get}|(%ctx, %obj, "context")
  |\hb{weakfill}|%1 = call |\api{@IsEmpty}|(%ctx)|\he|
  br %1, %end, %L1
L1:
  |\hb{weakfill}|%2 = call |\api{@IsObject}|(%ctx)|\he|
  br %2, %L3, %end
L3:
  |\hb{badfill}|%4 = call |\api{@Unwrap\textless Context2d\textgreater}|(%ctx)|\he|
  ...
end:
  ret
}
\end{minted}
\end{minipage}\end{lrbox}%
\begin{lrbox}{\lcGraph}\begin{tikzpicture}[x=1cm, y=1cm]
\node[vfarg]   (this)   at (2.05, 0)     {This};
\node[vf]      (as)     at (2.05, -0.8)  {As\textless Object\textgreater};
\node[vf]      (formal) at (2.05, -1.6)  {\%This};
\node[vfstart] (get)    at (2.05, -2.4)  {Get};
\node[vftest]  (empty)  at (0.42, -2.4)  {IsEmpty};
\node[vftest]  (isobj)  at (1.60, -3.3)  {IsObject};
\node[vfuse]   (unwrap) at (2.87, -3.3)  {Unwrap};
\node[vf]      (br1)    at (0.42, -3.3)  {br};
\node[vf]      (br2)    at (1.60, -4.0)  {br};
\draw[vfreg]  (this) -- (as);
\draw[vfcall] (as) -- (formal);
\draw[vfmem]  (formal) -- node[vflab, right=1pt] {\%obj} (get);
\draw[vfmem]  (get) -- node[vflab, above=0.5pt] {\%ctx} (empty);
\draw[vfmem]  (get) -- node[vflab, pos=0.55] {\%ctx} (isobj);
\draw[vfmem]  (get) -- node[vflab, pos=0.55] {\%ctx} (unwrap);
\draw[vfreg]  (empty) -- (br1);
\draw[vfreg]  (isobj) -- (br2);
\vforder{this}{1}\vforder{as}{2}\vforder{formal}{3}
\vforder{get}{4}\vforder{empty}{5}\vforder{br1}{6}\vforder{isobj}{7}
\vforder{br2}{8}\vforder{unwrap}{9}
\end{tikzpicture}\end{lrbox}%
\begin{lrbox}{\lcXlir}\footnotesize
\newcommand{\xhl}[2]{\begingroup\setlength{\fboxsep}{1pt}\colorbox{#1}{$#2$}\endgroup}%
$\begin{array}[t]{@{}r@{\;\;}l@{}}
&\xhl{entryfill}{\kw{fun}\ \mathit{SetWidth}()=}\\
\vfo{1}&\quad\xhl{argfill}{\Let{x}{\Arg{0}}};\\
&\quad\mathit{resurface}(x);\ \kw{return}\\
\vfo{3}&\kw{fun}\ \mathit{resurface}(o)=\\
\vfo{4}&\quad\proj{\neg\tnull\wedge\neg\tundef}{o};\\
\vfo{4}&\quad\Let{c}{\Host{\Any}{o}};\\
\vfo{6}&\quad\kw{if}\ \xhl{weakfill}{\ischk{\None}{c}}\ \kw{then}\ \eseq\ \kw{else}\\
\vfo{8}&\qquad(\kw{if}\ \xhl{weakfill}{\ischk{\AnyObj}{c}}\ \kw{then}\\
\vfo{9}&\qquad\quad\xhl{badfill}{\proj{\kw{Context2d}}{c}}\\
&\qquad\ \kw{else}\ \eseq);\\
&\quad\kw{return}
\end{array}$\end{lrbox}%
%
\setlength{\lcHeight}{\dimexpr\ht\lcListing+\dp\lcListing\relax}%
\ifdim\lcHeight<\dimexpr\ht\lcGraph+\dp\lcGraph\relax
  \setlength{\lcHeight}{\dimexpr\ht\lcGraph+\dp\lcGraph\relax}\fi
\ifdim\lcHeight<\dimexpr\ht\lcXlir+\dp\lcXlir\relax
  \setlength{\lcHeight}{\dimexpr\ht\lcXlir+\dp\lcXlir\relax}\fi
\addtolength{\lcHeight}{10pt}%
\begin{minipage}[t][\lcHeight][t]{\lcLeft}
\vspace{0pt}
\centering
\usebox{\lcListing}
\vfill
\subcaption{LLVM IR.}\label{fig:lowering-canvas-ir}
\end{minipage}%
\begin{minipage}[t][\lcHeight][t]{\lcMid}
\vspace{0pt}
\centering
\usebox{\lcGraph}
\vfill
\subcaption{SVFG.}\label{fig:lowering-canvas-svfg}
\end{minipage}%
\begin{minipage}[t][\lcHeight][t]{\lcRight}
\vspace{0pt}
\centering
\usebox{\lcXlir}
\vfill
\subcaption{XLIR program.}\label{fig:lowering-canvas-xlir}
\end{minipage}
\vspace{-8pt}
\caption{
The simplified LLVM IR of Figure~\ref{fig:motivating-a}
without the fix,
its SVFG,
and its XLIR program.
Colors mark the entry point (grey),
the argument read (violet),
the tests (orange),
and the use (red).
In~(b), dashed edges pass through a stack slot
and the dotted edge through a call.
}
\label{fig:lowering-canvas}
\vspace{-6pt}
\end{figure}

\point{Boundary value-flow tracking}
The SVFG tracks how values flow between LLVM operands,
but XLIR statements refer to
variables that stand for boundary values.
Lowering therefore maintains a map~$M$
from each LLVM operand that holds a boundary value,
and each memory location \code{*\%p}
that a pointer operand \code{\%p} points to,
to the XLIR variable that stands for this value.
An operand that holds the result of a type check
is mapped to the outcome~$\ischk{\tau}{v}$ of the check,
where~$v$ is the checked variable.
Thus,
$M$ maps all copies of a boundary value
in registers and memory
to a single XLIR variable,
and carries the outcome of a type check
to the branch that tests it.
For example,
the operand \code{\%ctx} of \code{Unwrap}
(line~16, Figure~\ref{fig:lowering-canvas-ir})
points to a stack slot,
which $M$ maps to the variable~$c$
for the value that \code{Get} writes there (line~9).
Hence,
the call becomes $\proj{\kw{Context2d}}{c}$
(Figure~\ref{fig:lowering-canvas-xlir}).

The lowering pass is a worklist algorithm over the SVFG.
The worklist starts with the instructions
that read the entry point's arguments
(e.g., line~2 of Figure~\ref{fig:lowering-canvas-ir}).
The algorithm repeatedly takes from the worklist
the instruction~$i$ that comes first in control-flow order,
updates~$M$ and translates~$i$ into XLIR
by the rule of Table~\ref{tab:lowering} for~$i$,
and adds the SVFG successors of~$i$ to the worklist.
For a call to the binding API,
the rule follows from the effects
that~$\mathcal{K}$ gives for the called function.
In Figure~\ref{fig:lowering-canvas},
the circled numbers give this order
and link each XLIR statement to its instruction.

Native functions can also exchange boundary values
through memory,
by passing pointers.
Since XLIR has no memory,
a value that a native function reads through a pointer
becomes an additional parameter,
and a value it writes through a pointer
becomes an additional result.
For example,
when a native helper writes a host operation's result
through an out-parameter,
the corresponding XLIR function returns that value,
and the caller binds it to a fresh variable.
We write $\mathit{in}(g)$ and $\mathit{out}(g)$
for the memory locations
that a function~$g$ may read and write,
which the SVFG identifies.

\begin{table}[t]
\caption{
Lowering rules per LLVM instruction~$i$:
when each applies,
what it adds to the map~$M$,
and its XLIR translation~$\sem{i}$
($\sem{\cdot}$ also translates blocks and function bodies).
$\vof{\%a}$ looks up~\code{\%a} in~$M$,
$\mathcal{K}$ is the binding API specification (Table~\ref{tab:spec}),
\code{\%a}$_j$ is the operand at position~$j$,
and $\mathit{in}(g)$ and $\mathit{out}(g)$ are the memory locations
that a function~$g$ of the module may read and write.
}
\label{tab:lowering}
\vspace{-8pt}
\footnotesize
\begin{tabular}{@{}l@{\quad}l@{\quad}l@{\quad}l@{}}
\toprule
\textbf{Instruction $i$} & \textbf{Condition} & \textbf{Added to $M$} & \textbf{XLIR} $\sem{i}$\\
\midrule
\code{\%y = \textbf{call} @f(\%a)} & $\langle\kw{hop},\tau,\code{ret}\rangle\in \mathcal{K}(f)$ & $\texttt{\%y} \mapsto y$ & $\Let{y}{\Host{\tau}{\vof{\%a}}}$\\
\code{\%y = \textbf{call} @f(...)} & $\langle\kw{is},\tau,j\rangle\in \mathcal{K}(f)$ & $\texttt{\%y} \mapsto \ischk{\tau}{\vof{\%a$_j$}}$ & $\eseq$\\
\code{\%y = \textbf{call} @f(...)} & $\langle\kw{use},\tau,j\rangle\in \mathcal{K}(f)$ & & $\proj{\tau}{\vof{\%a$_j$}}$\\
\code{\%y = \textbf{call} @g(\%a)} & $g$ in the slice & $(\texttt{\%y}, \mathit{out}(g)) \mapsto \bar{y}$ & $\Let{\bar{y}}{g(M(\texttt{\%a}, \mathit{in}(g)))}$\\
\code{\textbf{define} @g(\%p)} & $g$ in the slice & $(\texttt{\%p}, \mathit{in}(g)) \mapsto \bar{p}$ & $\kw{fun}\ g(\bar{p})=\sem{\text{body of }g}$\\
\code{\textbf{ret} \%a} & in $g$ & & $\kw{return}\ M(\texttt{\%a}, \mathit{out}(g))$\\
\code{\%y = \textbf{getelementptr} \%a, f} & $f$ a field of $\tau$'s struct & & $\proj{\tau}{\vof{\%a}}$\\
\code{\%y = \textbf{cast} \%a} & & $\texttt{\%y} \mapsto \vof{\%a}$ & $\eseq$\\
\code{\textbf{store} \%a, \%p} & & $\texttt{*\%p} \mapsto \vof{\%a}$ & $\eseq$\\
\code{\%y = \textbf{load} \%p} & & $\texttt{\%y} \mapsto \vof{*\%p}$ & $\eseq$\\
\code{\textbf{br} \%c, L1, L2} & $\vof{\%c} = \ischk{\tau}{v}$ & & $\Ifs{\ischk{\tau}{v}}{\sem{\texttt{L1}}}{\sem{\texttt{L2}}}$\\
\code{\textbf{br} \%c, L1, L2} & otherwise & & $\Ifs{\star}{\sem{\texttt{L1}}}{\sem{\texttt{L2}}}$\\
\bottomrule
\end{tabular}
\vspace{-6pt}
\end{table}

\point{Translation}
Lowering translates into an XLIR function
every function that contains an instruction of the slice,
along with the functions on the call paths
from the entry point to such a function.
Argument reads, host operations, uses, calls and returns
become the corresponding XLIR statements,
with~$M$ naming their operands,
and an API call with several effects
is translated once for each effect.
Type checks, casts, loads and stores
only update~$M$ and translate to~$\eseq$,
as do instructions outside the slice.
A conditional branch becomes an XLIR $\kw{if}$
that contains the translations of its two paths
up to the point where they merge again,
i.e., the branch's immediate post-dominator in the CFG.
Its condition is the check outcome that~$M$ gives for it,
if any,
and~$\star$ otherwise.

To better showcase the remaining two phases,
we switch to the Python motivating example
(\cref{fig:motivating-b}),
whose entry point \code{write} lowers
to the XLIR program of \cref{fig:xlir-ex}.

\subsection{Tag Refinement Analysis}
\label{sec:tag-refinement}
To detect boundary type errors
(Definition~\ref{def:bte}),
we compute the type of each variable at every program point
using a~\textit{context-sensitive forward} dataflow analysis
over the interprocedural control-flow graph (CFG)
of the XLIR program~$P$
produced by the previous phase
(Section~\ref{sec:lowering}).

\begin{figure}[t]
\captionsetup[sub]{font=footnotesize,skip=3pt}%
\begin{minipage}[b]{0.46\textwidth}
\centering\normalsize
$\begin{array}{@{}l@{\quad}l@{}}
\Let{x}{\Arg{i}} & \Gamma'=\Gamma[x\mapsto\neg\None]\\
\Let{y}{\Host{\tau}{x}} & \Gamma'=\Gamma[y\mapsto\tau]\\
\proj{\tau}{x} & \Gamma'=\Gamma[x\mapsto\Gamma(x)\wedge\tau]\\
\ischk{\tau}{x}\ \texttt{T} & \Gamma'=\Gamma[x\mapsto\Gamma(x)\wedge\tau]\\
\ischk{\tau}{x}\ \texttt{F} & \Gamma'=\Gamma[x\mapsto\Gamma(x)\wedge\neg\tau]\\
\star\ \texttt{T},\ \star\ \texttt{F} & \Gamma'=\Gamma\\
\Let{\bar y}{f(\bar x)} & \Gamma'=\Gamma[\bar y\mapsto\mathit{ret}_f]
\end{array}$
\vspace{1.5\baselineskip}
\subcaption{Transfer functions.}\label{fig:taginf-rules}
\par{\footnotesize\vspace{\baselineskip}}
\end{minipage}\hfill
\begin{minipage}[b]{0.5\textwidth}
\centering
\scalebox{0.68}{\input{figures/fig-taginf}}
\subcaption{
The CFG of the XLIR program of Figure~\ref{fig:xlir-ex} from line~3 on,
without the blue fix.}\label{fig:taginf-ex}
\end{minipage}
\vspace{-9pt}
\caption{Tag refinement.}
\label{fig:taginf}
\vspace{-6pt}
\end{figure}

\point{Domain and transfer functions}
The domain is the type environments $\Gamma$, which map each
variable to a type.
We start at the first statement with $\Gamma=\{\}$.
Each statement maps
the environment $\Gamma$ before it to $\Gamma'$
after it (Figure~\ref{fig:taginf-rules}).
A use refines the type of $x$ like a successful test,
because a run stops at an unsafe use (\RN{S-UseUnsafe}).
A call passes the types of its arguments to the parameters of~$f$
and gives $\bar y$ the types $\mathit{ret}_f$
that $f$ returns in that calling context
(\cref{app:calls}).
At a merge point,
a variable has the union~$\vee$ of its types
on the incoming edges.

\point{Example}
Figure~\ref{fig:taginf-ex} shows
the CFG of the XLIR program of \code{write}
from Figure~\ref{fig:motivating-b}
(\cref{fig:xlir-ex}),
starting at line~3,
with the types $\Gamma(b)$ on its edges.
The host operation on line~2 (not shown)
gives $b$ type $\Any$,
and the tests refine its type
along their branches.
The else-branch of the choice on line~4 is empty,
leaving $\neg\None$ at the join.
At the use on line~8, $\Gamma(b)$ is $\neg\None\wedge\tstr$.


\point{Soundness and precision}
Our transfer functions over-approximate the rules of the semantics (\cref{app:semantics}), and
joins compute unions.
As such,
in every run of $P$,
the actual value of $x$
before each statement belongs to $\sem{\Gamma(x)}$.
This set may also contain
values that $x$ never takes,
in part because the analysis is non-relational,
so joins lose correlations between variables.
To limit such imprecision across calls,
the analysis is context-sensitive:
it distinguishes each call by its $k$ most recent call sites
($k$-CFA),
so facts from different callers are not merged.

\point{Termination}
Recursive calls may introduce cycles in the CFG, so we iterate
to a fixpoint, joining each new environment with the previous
one until nothing changes up to type equivalence. A join can only enlarge the
type of a variable, and every
type that the analysis builds is a Boolean combination of the $n$ types (counting $\None$) that
appear as labels of $\kw{is}$, $\kw{hop}$, and $\kw{use}$ in $P$. Up to
equivalence, there are at most $2^{2^{n}}$ such combinations, since each is
determined by its truth table over these $n$ types. With finitely many
contexts of at most $k$ call sites, the iteration thus terminates.


\subsection{Solver-Based Subtyping Check}
\label{sec:smt}
\begin{figure}[t]
\renewcommand{\theFancyVerbLine}{\textcolor{black!40}{\sffamily\fontsize{5.5}{6.5}\selectfont\arabic{FancyVerbLine}}}%
\setminted{fontsize=\fontsize{6.5}{8.5}\selectfont, xleftmargin=10pt, numbersep=4pt, escapeinside=@@}%
\newcommand{\smtc}[1]{\textcolor{black!45}{; #1}}%
\captionsetup[sub]{justification=centering,singlelinecheck=false,font=footnotesize,position=below,skip=3pt}%
\begin{minipage}[t]{0.34\textwidth}
\vspace{0pt}
\begin{minted}[linenos]{text}
(declare-sort Tag 0)
(declare-datatype Val
  ((failed) (tagged (tag Tag))))
(declare-fun sub (Tag Tag) Bool)
(define-fun In ((v Val) (t Tag)) Bool
  (and (not (= v failed))
       (or (= (tag v) t)
           (sub (tag v) t))))
\end{minted}
\end{minipage}\hfill
\begin{minipage}[t]{0.33\textwidth}
\vspace{0pt}
\begin{minted}[linenos]{text}
(declare-const Str Tag)
(declare-const AsciiStr Tag)
(declare-const Bytes Tag)
(define-fun CPython ((v Val)) Bool
  (and (=> (In v AsciiStr)
           (In v Str))
       (=> (= v (tagged Str))
           (not (In v AsciiStr)))
       (not (and (In v Str)
                 (In v Bytes)))))
\end{minted}
\end{minipage}\hfill
\begin{minipage}[t]{0.31\textwidth}
\vspace{0pt}
\begin{minted}[linenos]{text}
(declare-const b Val)
(assert (CPython b))
(assert (not (= b failed)))
(assert (In b Str))
@\hb{badfill}@(assert (not (In b AsciiStr)))@\he@
(check-sat)             @\textcolor{baddraw}{; sat}@
@\fixl\hb{chkfill!60}@(assert (or (In b Bytes)@\hc@
            (In b AsciiStr)))@\he@
(check-sat)             @\textcolor{chkdraw}{; unsat}@
\end{minted}
\end{minipage}\par
\begin{minipage}[t]{0.34\textwidth}
\subcaption{Values and types.}\label{fig:smt-io-values}
\end{minipage}\hfill
\begin{minipage}[t]{0.33\textwidth}
\subcaption{The tag hierarchy.}\label{fig:smt-io-order}
\end{minipage}\hfill
\begin{minipage}[t]{0.31\textwidth}
\subcaption{The query.}\label{fig:smt-io-query}
\end{minipage}
\vspace{-8pt}
\caption{The subtyping check for the use
$\proj{\AsciiStr}{b}$ of \code{write}
(line~8 of Figure~\ref{fig:xlir-ex}).
Red marks the assertion
that~$b$ is not an \code{AsciiStr}
(line~5 of~(c)).
Blue marks what the fix of Figure~\ref{fig:motivating-b}
adds (lines~7--8).
}
\label{fig:smt-io}
\vspace{-6pt}
\end{figure}

By Definition~\ref{def:bte},
a use $\proj{\tau}{x}$ is a boundary type error
if some run reaches it with a value of~$x$
outside~$\llbracket\tau\rrbracket$.
Tag refinement guarantees that every value~$x$ may take at the use
lies in~$\llbracket\Gamma(x)\rrbracket$ (Section~\ref{sec:tag-refinement}).
Hence,
the use is safe if $\Gamma(x)$ is a subtype of~$\tau$,
i.e., if $\llbracket\Gamma(x)\rrbracket \subseteq \llbracket\tau\rrbracket$.
We decide this inclusion by refutation:
we search for a counterexample,
\emph{a value that has type~$\Gamma(x)$ but not type~$\tau$}:
\begin{equation}
\exists v.\; v \in \llbracket\Gamma(x)\rrbracket
\wedge v \notin \llbracket\tau\rrbracket,
\label{eq:counterexample}
\end{equation}
or equivalently,
$\llbracket\Gamma(x) \wedge \neg\tau\rrbracket \neq \emptyset$.
If no such value exists,
the inclusion holds,
and the use is safe.
Otherwise,
the value passes every test preceding
the use but violates the use's assumption.
Therefore,
we report the use
with this value serving as a~\emph{witness}.

Since XLIR types denote sets of values~\cite{FCB08}
and are Boolean combinations of tags,
the above question reduces to satisfiability.
Each type translates to a logical formula over a value~$v$,
and a value of type $\Gamma(x) \wedge \neg\tau$ exists
exactly when the formula for this type
is satisfiable under the constraints of the tag hierarchy.
Following Dminor~\cite{BGHL10},
we pose this question to an SMT solver.
We explain our encoding through the example
of Figure~\ref{fig:smt-io},
which shows the query
for $\proj{\AsciiStr}{b}$
of Figure~\ref{fig:taginf-ex}.

\point{Values and types (Figure~\ref{fig:smt-io-values})}
The encoding follows the denotations of
Figure~\ref{fig:types}.
A value is $\emp$ or a tag, and \code{(In v t)} states
that a value \code{v} has type $t$,
where \code{sub} is the order~$\preceq$.
\code{Tag} is an uninterpreted sort,
because our encoding treats
the set of tags as~\emph{open}:
a value may carry a tag that $P$ never names,
such as a subclass of \code{str} that a script defines.

\point{Tag hierarchy (Figure~\ref{fig:smt-io-order})}
Whether a value of type $\Gamma(x) \wedge \neg\tau$ exists
depends on the tag hierarchy~$\mathcal{T}$,
of which the solver knows nothing
unless the query states it.
Each query therefore includes the axioms of~$\mathcal{T}$,
which the frontend derives from the host's built-in tags
and the tags the extension registers
(Section~\ref{sec:entry-point}).
For CPython,
these include that
\code{AsciiStr} lies below \code{Str} (lines~5--6),
that \code{Str} does not lie below \code{AsciiStr} (lines~7--8),
and that no value is both a \code{Str} and a \code{Bytes}
(lines~9--10).
Rather than stating these axioms for all values,
which would require quantifiers,
the query applies them only to the value it asks about
(Figure~\ref{fig:smt-io-query}, line~2).
This suffices,
since the query concerns that value alone,
and keeps it quantifier-free and thus decidable.

\point{SMT query (Figure~\ref{fig:smt-io-query})}
The query encodes Equation~\eqref{eq:counterexample}.
It declares a value for $x$ and asserts three things about it:
first, the tag hierarchy facts (line~2),
second, that it has type $\Gamma(x)$ (lines~3--4),
and third, that it \emph{does not} have type $\tau$ (line~5). If the solver answers
unsat, the use is safe. Otherwise, \tool\ reports the use as type-unsafe, with the
solver's model as a witness. For $\proj{\AsciiStr}{b}$, the solver answers sat (line~6),
and its model gives $b$ the tag $\tstr$, which
is the tag of a string that is not ASCII, such as the one that the
reproducer of Figure~\ref{fig:motivating-b} writes.
With the fix,
$\Gamma(b)$ is equivalent to $\neg\None\wedge(\tbytes\tor\AsciiStr)\wedge\tstr$,
which adds lines~7--8 to the query.
Since $b$ is not an $\AsciiStr$ (line~5), it is a $\tbytes$,
and no value is both a $\tbytes$ and a $\tstr$,
so the query is unsat (line~9).

\subsection{Implementation Details and Discussion}
\label{sec:implementation}

We implement our approach in \tool,
a command-line tool of roughly 20K lines of C++
built on LLVM~21.1.0,
with Z3~\cite{z3} as the SMT solver.
It builds on SVF~\cite{svf},
a static value-flow analysis framework for LLVM,
which computes the SVFG and the CFG
that lowering uses (\cref{sec:lowering}).
Tag refinement uses a context depth of $k=2$
(\cref{sec:tag-refinement}).
For each type-unsafe use,
\tool\ reports the boundary value,
the tag the use assumes,
and a witness (\cref{sec:smt}).
Since~\tool\ checks each use separately,
a single bug (i.e., a missing or incomplete test)
may produce multiple reports,
e.g., an unchecked argument with several uses.

\point{Generalizability and extensibility}
\tool\ supports extensions written against
N-API, NAN and {\tt node-addon-api} for JavaScript,
and against the CPython C-API for Python.
A new binding framework requires
rules that identify its entry points
(Section~\ref{sec:entry-point})
and a binding API specification
(Section~\ref{sec:lowering}).
A new language additionally requires
the tag hierarchy of its host
and a way to print witnesses in the language.
We plan to support {\tt pybind11} and Cython.

\point{Limitations}
\tool\ targets only boundary type errors,
as defined in Definition~\ref{def:bte}.
It does not detect errors that depend on a value's contents,
such as
(1)~an empty list where native code expects a non-empty one,
or (2)~reference-counting errors~\cite{MYZYZ23}.
Its effectiveness also depends on three factors.
First,
it relies on binding API specifications
given as input for each binding API
(354 entries for the JavaScript APIs and 280 for CPython).
An incorrect or missing entry
can lead to false positives or false negatives.
Second,
it depends on entry-point identification
(Section~\ref{sec:entry-point}).
\tool\ finds entry points through their registrations,
so it misses those that a module registers dynamically,
rather than during its initialization
(i.e., loading time).
However,
prior work has shown
that dynamic registration is rare in practice~\cite{pyxray,gasket}.
Third,
it is limited by the code it can analyze.
SVF does not analyze asynchronous code,
or code invoked by C libraries linked dynamically.
\Cref{app:fn-fp} discusses
examples of false positives and false negatives.

\section{Evaluation}
\label{sec:evaluation}

We evaluate~\tool{} based on
the following research questions:

\begin{enumerate}[label={\bf RQ\arabic*}, leftmargin=2.3\parindent]
\item Can~\tool\ detect
previously known type errors in native extensions?
What is its precision and recall?
(Section~\ref{sec:rq1-effectiveness})
\item Can~\tool\ find
new boundary type errors in real-world native extensions?
(Section~\ref{sec:rq2-bug-finding-results})
\item What are the characteristics of
the type errors discovered by~\tool?
(Section~\ref{sec:rq3-bug-characteristics})
\item What is the performance of~\tool?
(Section~\ref{sec:rq4-performance})
\end{enumerate}

\point{Hardware}
A 16-core 3.6GHz machine
with 32GB of RAM, running Ubuntu 22.04.

\subsection{RQ1: Precision and Recall of~\tool}
\label{sec:rq1-effectiveness}

\subsubsection{Experimental Setup}
\label{sec:rq1-setup}

We want to assess
whether~\tool{} detects
previously known boundary type errors,
and how many of its reported type errors
are genuine bugs.

\point{JavaScript benchmarks}
We use two JavaScript benchmark suites
from the study of~\citet{flowjs}.
Their~\textit{micro-benchmark} suite consists
of~\empirical{25} small native extensions,
each with a single C/C++ function
extracted from a real-world extension,
across different binding frameworks
(e.g., \code{NAN} and
{\tt node-addon-api}).
It contains~\empirical{30} type errors,
each with a reproducer.

The~\textit{macro-benchmark}
suite incorporates the
33 real-world npm packages
in which~\citet{flowjs} identified
type errors.
For each type error that the study flagged
in a native entry function,
we attempt to construct a reproducer
that passes a wrongly-typed value
to the function
and triggers the error,
and we keep the error only if we succeed.
We exclude~\empirical{seven} packages,
because the buggy version is no longer
available on npm
or because we could not reproduce
any of its errors.
The final suite contains~\empirical{26} packages,
each pinned to its buggy version,
with~\empirical{72} type errors.

\point{Python benchmarks}
No ground truth exists for Python,
so we build one from GitHub.
We search for issues and pull requests at
the Python-C boundary in
the \nnum{1000} most downloaded PyPI packages
whose native extensions
are written against the raw CPython C-API.
We keep an issue if
a maintainer has acknowledged and fixed it
and it is in scope,
i.e.,
a native function receives an argument
of the wrong type,
or a call back into Python
(e.g., {\tt \_\_len\_\_})
returns a value of the wrong type.
The final suite contains~\empirical{42}
boundary type errors
from~\empirical{21} GitHub issues
of~\empirical{13} packages.

\begin{table}[t]
\centering
\small
\setlength{\tabcolsep}{2.5pt}
\setlength{\aboverulesep}{0pt}\setlength{\belowrulesep}{0pt}%
\renewcommand{\grouprule}{\noalign{\begingroup\color{black!25}\hrule height 0.3pt\endgroup}}%
\renewcommand{\arraystretch}{1.25}%
\caption{
Results of~\tool\ on 26 npm packages.
``CVE'' lists the package's known CVEs,
if any,
and ``Time'' is the analysis time of~\tool\ in seconds.
``Sites detected'' is the number
of~\emph{known} type errors
that~\tool\ finds,
while TP and FP split the type errors
it reports
into true positives and false positives.
}
\label{tab:macro-suite}
\vspace{-7pt}
\resizebox{\textwidth}{!}{
\begin{tabular}{@{}lllrcrr@{\hspace{0.6em}{\color{black!30}\vrule width 0.4pt}\hspace{0.6em}}lllrcrr@{}}
\toprule
 &  &  & \textbf{Time} & \textbf{Sites} &  &  &  &  &  & \textbf{Time} & \textbf{Sites} &  &  \\
\textbf{Package} & \textbf{Version} & \textbf{CVE} & \textbf{(s)} & \textbf{detected} & \textbf{TP} & \textbf{FP} & \textbf{Package} & \textbf{Version} & \textbf{CVE} & \textbf{(s)} & \textbf{detected} & \textbf{TP} & \textbf{FP} \\
\midrule
\texttt{utf-8-validate} & 5.0.8 & -- & 0.1 & 2/2 & 1 & 0 & \texttt{fast-string-search} & 1.4.1 & {\renewcommand{\arraystretch}{1}\begin{tabular}[c]{@{}l@{}}\href{https://nvd.nist.gov/vuln/detail/CVE-2022-22138}{2022-22138}\\\href{https://nvd.nist.gov/vuln/detail/CVE-2022-25872}{2022-25872}\end{tabular}} & 0.1 & 2/2 & 18 & 0 \\
\grouprule
\texttt{sqlite3} & 5.0.1 & {\renewcommand{\arraystretch}{1}\begin{tabular}[c]{@{}l@{}}\href{https://nvd.nist.gov/vuln/detail/CVE-2022-21227}{2022-21227}\\\href{https://nvd.nist.gov/vuln/detail/CVE-2022-43441}{2022-43441}\end{tabular}} & 10.8 & 6/6 & 13 & 2 & \texttt{gs-node-lmdb} & 0.9.0 & -- & 3.9 & 6/6 & 47 & 1 \\
\grouprule
\texttt{bigint-buffer} & 1.1.5 & \href{https://nvd.nist.gov/vuln/detail/CVE-2025-3194}{2025-3194} & 0.1 & 1/1 & 3 & 0 & \texttt{sendto} & 1.0.3 & -- & 0.1 & 2/2 & 3 & 0 \\
\grouprule
\texttt{libpq} & 1.8.9 & \href{https://nvd.nist.gov/vuln/detail/CVE-2022-25852}{2022-25852} & 0.8 & 3/3 & 28 & 6 & \texttt{rapid-crc} & 1.0.10 & -- & 0.1 & 1/1 & 3 & 3 \\
\grouprule
\texttt{libxmljs} & 0.19.7 & \href{https://nvd.nist.gov/vuln/detail/CVE-2022-21144}{2022-21144} & 186.1 & 5/5 & 83 & 7 & \texttt{csocket} & 1.0.3 & -- & 0.3 & 3/3 & 19 & 0 \\
\grouprule
\texttt{int64-napi} & 1.0.1 & -- & 1.5 & 2/2 & 2 & 0 & \texttt{bkjs-utils} & 0.2.8 & -- & 4.2 & 1/1 & 25 & 1 \\
\grouprule
\texttt{bignum} & 0.13.1 & \href{https://nvd.nist.gov/vuln/detail/CVE-2022-25324}{2022-25324} & 0.9 & 3/3 & 27 & 1 & \texttt{csac-ed25519} & 0.0.3 & -- & 0.7 & 3/3 & 9 & 15 \\
\grouprule
\texttt{node-crc} & 1.3.0 & -- & 0.1 & 1/1 & 20 & 0 & \texttt{node-lzma} & 0.1.0 & -- & 1.2 & 1/1 & 4 & 0 \\
\grouprule
\texttt{node-mbus} & 1.2.1 & -- & 1.0 & 4/4 & 8 & 0 & \texttt{zopfli-node} & 2.0.3 & -- & 0.8 & 7/7 & 15 & 3 \\
\grouprule
\texttt{termios-fixedv12} & 0.1.9 & -- & 0.4 & 1/1 & 12 & 0 & \texttt{libasar\_enc} & 1.0.0 & -- & 0.1 & 2/2 & 3 & 0 \\
\grouprule
\texttt{jitterbuffer} & 0.1.14 & -- & 1.1 & 5/5 & 7 & 0 & \texttt{binary-diff} & 1.0.0 & -- & 0.7 & 2/2 & 4 & 0 \\
\grouprule
\texttt{multi-hashing} & 1.0.0 & -- & 12.9 & 1/1 & 39 & 2 & \texttt{@alien.sh/signals} & 1.0.0 & -- & 0.1 & 1/1 & 1 & 0 \\
\grouprule
\texttt{bigint-hash} & 0.2.2 & -- & 0.1 & 3/3 & 9 & 0 & \texttt{cuckaroo29b-hashing} & 1.0.0 & -- & 0.3 & 4/4 & 6 & 0 \\
\midrule
 &  &  &  &  &  &  & \textbf{Total} &  &  & \textbf{228.4} & \textbf{72/72} & \textbf{409} & \textbf{41} \\
\bottomrule
\end{tabular}}
\vspace{-6pt}
\end{table}

\point{Metrics}
For each suite,
we check whether~\tool\ detects
every type error in the ground truth.
Since \tool\ analyzes every entry point of a module,
it also reports type errors beyond the ground truth,
either previously unknown or false positives.
We validate each of them
with an LLM agent
(Claude Opus 5.5 with high reasoning effort),
which tries to craft a reproducer,
i.e.,
a program that invokes the native extension
with a wrongly-typed input
that reaches
the reported type-unsafe operation
and causes unexpected runtime behavior.
We~\emph{conservatively} count type errors
without a reproducer as false positives,
so precision is the fraction
of confirmed type errors.

\point{Baselines}
Two prior tools target
type errors in JavaScript native extensions:
FlowJS~\cite{flowjs}
and the checker of~\citet{brown-checker}.
Neither is publicly available,
so we cannot run them as baselines.
Section~\ref{sec:insights} explains
why both would miss type errors
that originate from ineffective checks
or values obtained through host operations.
Instead,
we run \tool\ on the macro-benchmark suite
and measure its~\emph{relative recall},
i.e., the fraction of its known errors that \tool\ reports.

For Python,
the only public tool
is MOPSA~\cite{mopsa},
which targets general runtime errors
and needs a hand-written driver per extension.
On 100 random extensions from our dataset,
it analyzes only 25,
and none of its alarms concerns a boundary type error,
so we do not use it as a baseline.

\subsubsection{Results}
\label{sec:rq1-results}

\point{JavaScript}
Table~\ref{tab:macro-suite} summarizes
the results of~\tool\
on the 26 packages of the macro-benchmark suite.
\tool\ analyzes all packages
in less than~\empirical{four} minutes
(i.e., \empirical{228.4} s)
and detects~\emph{all}
\empirical{72} known type errors
in the ground truth
(column ``Sites detected''),
achieving~\empirical{100\%} relative recall.
Overall,
\tool\ reports~\empirical{450} type errors,
of which~\empirical{409} are confirmed
by a reproducer.
This yields a precision of~\empirical{90.9\%}.
Beyond the~\empirical{72} type errors in the ground truth,
\tool\ confirms~\empirical{337} more in the same packages,
which FlowJS~\cite{flowjs} did not report.
We attribute these misses to two limitations of FlowJS.
First,
it operates on source code
and therefore does not see code generated by macros.
Second,
its analysis is intra-procedural.

On the micro-benchmark suite,
\tool\ detects~\empirical{28} of the~\empirical{30}
known type errors,
achieving~\empirical{93.3\%} recall.
Both misses are due to asynchronous code
(Section~\ref{sec:implementation},
``Limitations'').
\tool\ also reports~\empirical{seven}
false positives,
yielding a precision
of~\empirical{80\%}.

\point{Python}
Table~\ref{tab:python-results} summarizes
the results of~\tool\
on the~\empirical{21} known issues
from~\empirical{13} PyPI packages.
\tool\ analyzes all extension modules
in less than~\empirical{three} minutes
(i.e., \empirical{134.9}~s)
and detects~\empirical{33} of the~\empirical{42}
known type errors
(column ``Sites''),
achieving~\empirical{78.6\%} relative recall.
Of the~\empirical{nine} known errors that~\tool{} misses,
\empirical{six} are in {\tt atom},
which registers some of its native entry functions
through a runtime table
that \tool\ cannot identify
as a list of entry points.
Three involve a value returned through
an unmodeled CPython API call.
Overall,
\tool\ reports~\empirical{367} type errors,
i.e., \empirical{334} beyond the
\empirical{33} known ones it detects.
Of the~\empirical{367} type errors,
\empirical{323} are genuine
and verified by a reproducer.
This yields a precision of~\empirical{88.0\%}.
Notably,
many of the verified type errors
beyond the ground truth
exist in the latest releases of these packages.
We reported them to the maintainers,
who have acknowledged and fixed them
(Section~\ref{sec:rq2-bug-finding-results}).

\WFclear
\subsection{RQ2: Bug-Finding Results in the Wild}
\label{sec:rq2-bug-finding-results}

\begin{figure}[t]
\begin{minipage}[t]{0.535\textwidth}
\vspace{0pt}
\centering
\small\setlength{\tabcolsep}{3pt}
\captionof{table}{Results of~\tool\ on~\empirical{21}
known Python issues.
Column meanings are the same as in Table~\ref{tab:macro-suite}.}
\label{tab:python-results}
\scalebox{0.75}{%
\begin{tabular}{@{}llllcrrr@{}}
\toprule
\textbf{Package} & \textbf{Version} & \textbf{Issue} & \textbf{Module} & \textbf{Sites} & \textbf{T(s)} & \textbf{TP} & \textbf{FP} \\
\midrule
\texttt{markupsafe} & 1.1.0 & \href{https://github.com/pallets/markupsafe/issues/108}{\#108} & \texttt{\_speedups} & 1/1 & 0.1 & 1 & 0 \\
\grouprule
 &  & \href{https://github.com/pandas-dev/pandas/issues/66356}{\#66356} &  & 3/5 &  &  &  \\
\multirow{-2}{*}{\texttt{pandas}} & \multirow{-2}{*}{1.4.1} & \href{https://github.com/pandas-dev/pandas/issues/66797}{\#66797} & \multirow{-2}{*}{\texttt{\_libs.\_ujson}} & 1/1 & \multirow{-2}{*}{58.0} & \multirow{-2}{*}{49} & \multirow{-2}{*}{9} \\
\grouprule
 &  & \href{https://github.com/python-pillow/Pillow/issues/7913}{\#7913} & \texttt{\_imagingcms} & 1/1 & 0.2 & 1 & 0 \\
 &  & \href{https://github.com/python-pillow/Pillow/issues/9494}{\#9494} &  & 3/3 &  &  &  \\
\multirow{-3}{*}{\texttt{Pillow}} & \multirow{-3}{*}{10.2.0} & \href{https://github.com/python-pillow/Pillow/issues/9917}{\#9917} & \multirow{-2}{*}{\texttt{\_imaging}} & 1/1 & \multirow{-2}{*}{12.1} & \multirow{-2}{*}{10} & \multirow{-2}{*}{1} \\
\grouprule
\texttt{wrapt} & 2.1.2 & \href{https://github.com/GrahamDumpleton/wrapt/issues/326}{\#326} & \texttt{\_wrappers} & 5/5 & 1.7 & 149 & 4 \\
\grouprule
\texttt{zstandard} & 0.14.1 & \href{https://github.com/indygreg/python-zstandard/issues/102}{\#102} & \texttt{zstd} & 1/1 & 28.3 & 14 & 4 \\
\grouprule
 &  & \href{https://github.com/simplejson/simplejson/issues/284}{\#284} &  & 1/1 &  &  &  \\
\multirow{-2}{*}{\texttt{simplejson}} & \multirow{-2}{*}{3.17.4} & \href{https://github.com/simplejson/simplejson/issues/350}{\#350} & \multirow{-2}{*}{\texttt{\_speedups}} & 3/3 & \multirow{-2}{*}{9.5} & \multirow{-2}{*}{5} & \multirow{-2}{*}{9} \\
\grouprule
 &  & \href{https://github.com/ultrajson/ultrajson/issues/522}{\#522} &  & 1/1 &  &  &  \\
\multirow{-2}{*}{\texttt{ujson}} & \multirow{-2}{*}{5.2.0} & \href{https://github.com/ultrajson/ultrajson/issues/531}{\#531} & \multirow{-2}{*}{\texttt{ujson}} & 0/1 & \multirow{-2}{*}{3.8} & \multirow{-2}{*}{10} & \multirow{-2}{*}{3} \\
\grouprule
\texttt{mysqlclient} & 1.3.9 & \href{https://github.com/PyMySQL/mysqlclient/issues/137}{\#137} & \texttt{\_mysql} & 2/2 & 0.5 & 19 & 0 \\
\grouprule
\texttt{bitarray} & 1.8.1 & \href{https://github.com/ilanschnell/bitarray/issues/116}{\#116} & \texttt{\_bitarray} & 1/1 & 0.8 & 3 & 0 \\
\grouprule
 &  & \href{https://github.com/libgit2/pygit2/issues/1476}{\#1476} &  & 1/1 &  &  &  \\
\multirow{-2}{*}{\texttt{pygit2}} & \multirow{-2}{*}{1.19.3} & \href{https://github.com/libgit2/pygit2/issues/1478}{\#1478} & \multirow{-2}{*}{\texttt{\_pygit2}} & 1/1 & \multirow{-2}{*}{7.6} & \multirow{-2}{*}{34} & \multirow{-2}{*}{6} \\
\grouprule
\texttt{python-ldap} & 3.4.5 & \href{https://github.com/python-ldap/python-ldap/issues/556}{\#556} & \texttt{\_ldap} & 1/1 & 1.0 & 4 & 0 \\
\grouprule
\texttt{pygame-ce} & 2.4.1 & \href{https://github.com/pygame-community/pygame-ce/issues/3408}{\#3408} & \texttt{math} & 5/5 & 2.8 & 22 & 3 \\
\grouprule
 &  & \href{https://github.com/nucleic/atom/issues/256}{\#256} &  & 0/2 &  &  &  \\
 &  & \href{https://github.com/nucleic/atom/issues/258}{\#258} & \multirow{-2}{*}{\texttt{catom}} & 1/3 & \multirow{-2}{*}{8.4} & \multirow{-2}{*}{2} & \multirow{-2}{*}{5} \\
\multirow{-3}{*}{\texttt{atom}} & \multirow{-3}{*}{0.12.1} & \href{https://github.com/nucleic/atom/issues/259}{\#259} & \texttt{sortedmap} & 0/2 & 0.2 & 0 & 0 \\
\midrule
\textbf{Total} & & & & \textbf{33/42} & \textbf{134.9} & \textbf{323} & \textbf{44} \\
\bottomrule
\end{tabular}}
\end{minipage}\hfill
\begin{minipage}[t]{0.445\textwidth}
\vspace{0pt}
\centering
\includegraphics[width=\linewidth]{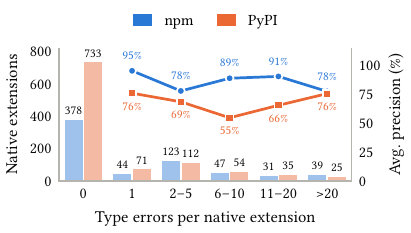}\par
\vspace{-12pt}
\captionof{figure}{Native extensions and their average precision
by type errors per extension.}
\label{fig:findings-precision}
\par\vspace{14pt}
\small
\captionof{table}{Reports
that we have filed so far for type errors
found by~\tool.
Our efforts are ongoing,
and more reports will follow.}
\label{tab:issue-status}
\vspace{-9pt}
\resizebox{0.75\linewidth}{!}{%
\renewcommand{\arraystretch}{1.05}%
\begin{tabular}{@{}lrrr@{}}
\toprule
\textbf{Status} & \textbf{JavaScript} & \textbf{Python} & \textbf{Total} \\
\midrule
Open & 13 & 19 & 32 \\
Fixed & 16 & 10 & 26 \\
Won't fix & 0 & 0 & 0 \\
\midrule
\textbf{Total} & \textbf{29} & \textbf{29} & \textbf{58} \\
Packages & 27 & 16 & 43 \\
\bottomrule
\end{tabular}}
\end{minipage}
\vspace{-10pt}
\end{figure}

We evaluate whether~\tool{} finds
\emph{unknown} boundary type errors in the wild.

\point{JavaScript benchmarks}
We take the~\nnum{1266} npm packages
with native extensions
of~\citet{gasket},
and add the~\empirical{100}
most popular~\cite{npm} npm packages
with native extensions that are
not among them.
Since~\tool\ takes LLVM bitcode as input,
we attempt to build,
for the~\emph{latest} release of each package,
a single bitcode file that links all of
its compilation units,
using the headers of Node.js 24.21.0.
Out of the~\empirical{\nnum{1366}} packages,
we successfully build LLVM bitcode
for~\empirical{\numNpmPackages}.
The remaining~\empirical{699} fail to build
because they (1) depend on outdated
V8 or Node.js headers,
(2) depend on system libraries unavailable
in our build environment,
or (3) use custom build scripts.

\point{Python benchmarks}
We start from the~\empirical{\nnum{100000}}
most downloaded PyPI projects,
ranked by monthly downloads according
to ecosyste.ms~\cite{ecosystem.ms},
and take their~\emph{latest} release.
We discard projects that are pure Python
(\empirical{\nnum{77162}})
or ship only binary wheels
with no source code information
(\empirical{\nnum{3586}}).
Of the remaining
\empirical{\nnum{19252}} packages,
we keep the~\empirical{\nnum{1571}}
whose source distributions
define at least one extension
module in C/C++ against the raw C-API.
Then,
we attempt to build
LLVM bitcode for each native extension,
and succeed for~\empirical{813} packages,
which together contain~\empirical{\numPyPIModules}
native extension modules.
The others fail to build,
mainly due to unavailable system libraries.

\point{Metrics}
Using a 30-minute timeout,
we run~\tool\ on each of the~\empirical{\numNpmPackages}
npm packages
and~\empirical{\numPyPIModules}
PyPI extension modules
and record its findings.
We validate each finding
with the LLM agent of RQ1
(Section~\ref{sec:rq1-effectiveness}),
and again count findings
without a reproducer as false positives.

\point{Results}
In total,
\tool\ reports~\empirical{\nnum{3909}} type errors
(counted per type-unsafe use, \cref{sec:implementation})
in~\empirical{284} of the~\empirical{\numNpmAnalyzed}
npm packages
it analyzes,
and~\empirical{\nnum{3011}} in~\empirical{297}
of the~\empirical{\numPyPIAnalyzed}
PyPI extension modules
(from~\empirical{251} packages).
Of these,
\empirical{\nnum{3286}} and~\empirical{\nnum{2244}}
are confirmed with a reproducer
(precision~\empirical{84\%} and~\empirical{75\%}).
Notably,
\tool\ achieves~\empirical{100\%} precision
in~\empirical{180} of the~\empirical{284} packages
and~\empirical{150} of the~\empirical{297} modules
with at least one report.
For~\empirical{5} npm packages
and~\empirical{22} PyPI modules,
the analysis does not complete within our timeout
(see Section~\ref{sec:rq4-performance}).
Precision is lower in PyPI
because CPython extensions detect failures
with plain C comparisons such as \code{== NULL},
often across several paths,
rather than with API calls such as V8's \code{IsEmpty()}.
\tool\ loses information
where such paths merge
(Section~\ref{sec:tag-refinement}),
which causes most of the PyPI false positives we inspected.


Figure~\ref{fig:findings-precision} breaks down
the native extensions by the number of type errors
\tool\ reports in each,
along with the average precision per group.
Most extensions with type errors have only a few:
\empirical{167} of the~\empirical{284} in npm
and~\empirical{183} of the~\empirical{297} in PyPI
have at most five.
Precision ranges from~\empirical{78\%}
to~\empirical{95\%} in npm
and from~\empirical{55\%} to~\empirical{76\%} in PyPI.

\point{Internal bindings}
We also run~\tool{} on the native modules
that ship with CPython~3.12
(\empirical{110} standard-library
and built-in extension modules)
and Node.js~24
(\empirical{71} internal bindings).
In CPython,
\tool\ finds~\empirical{94} genuine type errors
in~\empirical{10} modules,
including {\tt \_decimal},
{\tt \_ctypes},
{\tt \_tkinter},
{\tt \_ssl}
and {\tt \_io}.
In Node.js,
\tool\ finds one error reachable through a public API,
in {\tt vm.compileFunction},
which internally uses the buggy native module {\tt contextify}.

\begin{table}[t]
\caption{The most popular projects we filed reports for,
by their monthly downloads (ecosyste.ms~\cite{ecosystem.ms}).}
\label{tab:reported}
\vspace{-7pt}
\scriptsize\setlength{\tabcolsep}{3pt}
\centering
\begin{tabular}{@{}lrcl@{\hspace{10pt}}lrcl@{}}
\toprule
\multicolumn{4}{@{}l}{\textbf{JavaScript}} & \multicolumn{4}{l@{}}{\textbf{Python}}\\
\cmidrule(r{6pt}){1-4}\cmidrule{5-8}
\textbf{Package} & \textbf{Downloads} & \textbf{Fixed} & \textbf{Impact}
 & \textbf{Package} & \textbf{Downloads} & \textbf{Fixed} & \textbf{Impact} \\
\midrule
\href{https://github.com/nodejs/node}{\texttt{Node.js}} & -- & 0/1 & Hard crash & \href{https://github.com/python/cpython}{\texttt{CPython}} & -- & 3/9 & Crash, incorrect output \\
\href{https://www.npmjs.com/package/better-sqlite3}{\texttt{better-sqlite3}} & 55.6M & 1/1 & Hard crash & \href{https://pypi.org/project/protobuf/}{\texttt{protobuf}} & 609.5M & 0/1 & Hard crash \\
\href{https://www.npmjs.com/package/canvas}{\texttt{canvas}} & 40.7M & 1/1 & Hard crash & \href{https://pypi.org/project/pandas/}{\texttt{pandas}} & 546.9M & 0/1 & Incorrect output \\
\href{https://www.npmjs.com/package/lmdb}{\texttt{lmdb}} & 31.0M & 1/1 & Hard crash & \href{https://pypi.org/project/pillow/}{\texttt{Pillow}} & 436.5M & 1/2 & Crash, incorrect output \\
\href{https://www.npmjs.com/package/re2}{\texttt{re2}} & 20.3M & 1/1 & Hard crash & \href{https://pypi.org/project/wrapt/}{\texttt{wrapt}} & 389.0M & 1/1 & Hard crash \\
\href{https://www.npmjs.com/package/unix-dgram}{\texttt{unix-dgram}} & 17.5M & 0/1 & Hard crash & \href{https://pypi.org/project/zstandard/}{\texttt{zstandard}} & 153.7M & 0/1 & Hard crash \\
\href{https://www.npmjs.com/package/tree-sitter}{\texttt{tree-sitter}} & 15.3M & 0/1 & Hard crash & \href{https://pypi.org/project/psycopg2/}{\texttt{psycopg2}} & 63.8M & 0/1 & Hard crash \\
\href{https://www.npmjs.com/package/ttf2woff2}{\texttt{ttf2woff2}} & 2.0M & 0/1 & Hard crash & \href{https://pypi.org/project/mmh3/}{\texttt{mmh3}} & 56.3M & 0/1 & Hard crash \\
\href{https://www.npmjs.com/package/libpq}{\texttt{libpq}} & 1.2M & 0/1 & Hard crash & \href{https://pypi.org/project/tree-sitter/}{\texttt{tree-sitter}} & 47.6M & 1/1 & Hard crash \\
\href{https://www.npmjs.com/package/ibm_db}{\texttt{ibm\_db}} & 545.7K & 2/2 & Crash, side effects & \href{https://pypi.org/project/lazy-object-proxy/}{\texttt{lazy-object-proxy}} & 28.0M & 0/1 & Incorrect output \\
\bottomrule
\end{tabular}
\vspace{-6pt}
\end{table}
\point{Reporting issues}
We report the type errors
that~\tool\ finds in
CPython and Node.js,
and in~\empirical{27} npm
and~\empirical{16} PyPI packages,
which we select by prioritizing actively maintained projects.
So far,
we have filed~\empirical{58} GitHub issues
(Table~\ref{tab:issue-status}),
each explaining a type error and its impact,
and we continue to report more.
Before filing an issue,
we checked by hand that its reproducer
triggers the reported error.
Of the~\empirical{58} issues,
\empirical{26} have been fixed,
in~\empirical{21} packages and in CPython.
These include widely used ones,
such as CPython itself
(in its {\tt \_decimal} and {\tt \_tkinter} modules),
{\tt Pillow} (\empirical{436.5M} monthly downloads),
{\tt wrapt} (\empirical{389.0M}),
{\tt better-sqlite3} (\empirical{55.6M}),
{\tt canvas} (\empirical{40.7M})
and {\tt lmdb} (\empirical{31.0M})
from Table~\ref{tab:reported},
as well as~{\tt mysqlclient} (\empirical{16.3M}).
The rest are still open,
and none has been marked ``won’t fix''.
This shows that developers
consider these issues worth fixing.

\subsection{RQ3: Characteristics of Type Errors}
\label{sec:rq3-bug-characteristics}

We classify each of the~\empirical{\nnum{5530}} type errors
(\empirical{\nnum{3286}} in npm
and~\empirical{\nnum{2244}} in PyPI)
that RQ2 confirmed with a reproducer
(Section~\ref{sec:rq2-bug-finding-results})
along two dimensions.
The first is its~\textit{symptom},
i.e.,
its observable effect at run time.
The second is its~\textit{root cause}:
where native code obtains the
wrongly-typed value
(an argument or a host operation's result),
and whether that value is
unchecked or checked ineffectively.

\begin{figure}[t]
\begin{minipage}[b]{0.485\textwidth}
\centering
\includegraphics[scale=0.88]{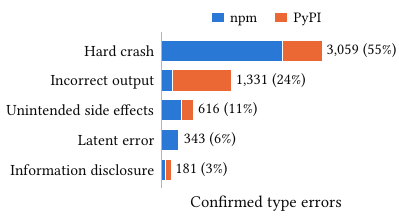}
\end{minipage}
\hfill
\begin{minipage}[b]{0.485\textwidth}
\centering
\includegraphics[scale=0.88]{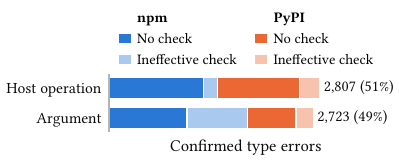}
\end{minipage}

\vspace{-18pt}
\begin{minipage}[t]{0.485\textwidth}
\captionof{figure}{Symptoms of the type errors
discovered by~\tool.}
\label{fig:symptoms}
\end{minipage}
\hfill
\begin{minipage}[t]{0.485\textwidth}
\captionof{figure}{Sources of the wrongly-typed values
in the type errors discovered by~\tool, by root cause.}
\label{fig:root-causes}
\end{minipage}
\vspace{-8pt}
\end{figure}

\point{Setup}
For the symptom,
we observe the runtime effect of the
reproducers generated in RQ2,
e.g., a crash (and its signal)
or a silently incorrect result.
For the root cause,
we use information
that~\tool\ records at every type-unsafe
operation it checks:
where the boundary value comes from,
and whether it is unchecked
or ineffectively checked.
Both steps are automated.

\point{Symptoms}
The most common symptom
(Figure~\ref{fig:symptoms})
is a~\textit{hard crash}
(\empirical{55\%}),
such as an assertion failure in the runtime
or a segmentation fault.
Hard crashes dominate in npm
(\empirical{70\%})
but not in PyPI (\empirical{34\%}),
where half of the type errors (\empirical{50\%})
cause~\textit{incorrect output}:
CPython's C-API
does not abort on a wrongly-typed value
but returns an error value (e.g., {\tt -1})
with an exception set,
and native code that ignores it
keeps executing.
In both ecosystems,
\empirical{11--12\%} of the type errors
cause~\textit{unintended side effects}:
after a failed host operation,
native code modifies program state or the environment
(e.g., a database).
Less frequent are~\textit{latent errors}
(\empirical{10\%} in npm,
below~\empirical{1\%} in PyPI),
where native code operates on an invalid value
with no observable effect,
and~\textit{information disclosure}
(\empirical{2\%} in npm,
\empirical{5\%} in PyPI),
where it returns
uninitialized memory or raw pointers.

\point{Root causes}
Figure~\ref{fig:root-causes} shows
the distribution of root causes,
grouped by the source of
the wrongly-typed boundary value.
In~\empirical{49\%} of the type errors
(\empirical{56\%} in npm,
\empirical{39\%} in PyPI),
the value is an argument
passed directly to the native entry function,
including its receiver
({\tt this} or {\tt self}).
In~\empirical{38\%} of these cases,
developers check the argument's type,
yet the check does not rule out
all unsafe types.
In the remaining~\empirical{51\%}
(\empirical{44\%} in npm,
\empirical{61\%} in PyPI),
the value is the result of a host operation:
a property or element read
({\tt canvas} in Figure~\ref{fig:motivating}),
or a call back into JavaScript or Python,
such as a user callback
or a special method like {\tt \_\_len\_\_}.
For these values,
developers perform no type check at all
in~\empirical{84\%} of the cases.
This indicates that
type checks in native extensions
tend to be shallow:
developers validate the arguments they receive,
but rarely the values nested inside them
or obtained through them.
\Cref{app:examples} discusses
three representative type errors in detail.

\subsection{RQ4: Performance of~\tool}
\label{sec:rq4-performance}

\begin{figure}[t]
\sbox{\plotbox}{\begin{minipage}[t]{0.5\textwidth}
\vspace{0pt}
\centering
\includegraphics{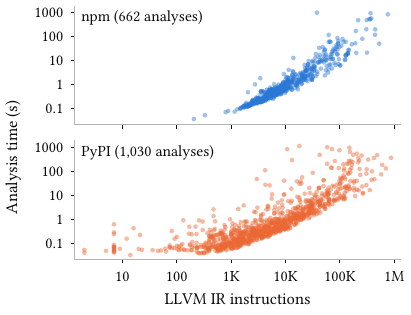}\par
\vspace{-12pt}
\captionof{figure}{Analysis time of~\tool{}
against the size of the analyzed LLVM IR,
for npm packages (top)
and PyPI extension modules (bottom).}
\label{fig:performance}
\end{minipage}}
\sbox{\tabcapbox}{\begin{minipage}[t]{0.485\textwidth}
\vspace{0pt}
\small
\captionof{table}{Time per analysis phase in seconds.
``Share'' is the fraction of the analysis time,
summed over all analyses,
spent in each phase,
and ``Other'' covers bitcode parsing,
reporting
and process teardown.
}
\label{tab:performance-percentiles}
\end{minipage}}
\begin{minipage}[t]{0.485\textwidth}
\vspace{0pt}
\usebox{\tabcapbox}\par
\vspace{-8pt}
\centering
\resizebox*{!}{0.72\linewidth}{%
\begin{tabular}[b]{@{}lrrrrr@{}}
\toprule
 & \multicolumn{4}{c}{\textbf{Time (s), by percentile}} & \\
\cmidrule(lr){2-5}
\textbf{Phase} & \textbf{50th} & \textbf{90th} & \textbf{99th} & \textbf{Max} & \textbf{Share} \\
\midrule
\multicolumn{6}{@{}l}{\textit{npm (662 analyses)}} \\
Lowering & 0.38 & 9.8 & 452.5 & 953.5 & 89.7\% \\
Tag refinement & 0.01 & 0.56 & 16.7 & 48.6 & 5.2\% \\
Subtyping check & 0.01 & 0.23 & 2.3 & 34.7 & 1.3\% \\
Other & 0.09 & 1.1 & 8.3 & 15.6 & 3.8\% \\
\cmidrule(l){1-6}
Total & 0.58 & 12.6 & 471.2 & 961.6 &  \\
\midrule
\multicolumn{6}{@{}l}{\textit{PyPI (1,030 analyses)}} \\
Lowering & 0.29 & 7.1 & 261.4 & 937.7 & 64.1\% \\
Tag refinement & 0.01 & 3.0 & 86.4 & 982.4 & 31.6\% \\
Subtyping check & 0.00 & 0.23 & 3.6 & 38.0 & 1.4\% \\
Other & 0.07 & 1.3 & 6.4 & 17.0 & 2.9\% \\
\cmidrule(l){1-6}
Total & 0.45 & 15.9 & 459.5 & \nnum{1115}.7 &  \\
\bottomrule
\end{tabular}}
\end{minipage}
\hfill
\usebox{\plotbox}
\vspace{-10pt}
\end{figure}

We measure
the performance of~\tool{}
and assess whether it is fast enough to run
routinely early in the development pipeline
(e.g., on every commit in continuous integration).

\point{Setup}
We run~\tool\ on the~\empirical{\numNpmPackages} npm packages
and the~\empirical{\numPyPIModules} PyPI extension modules
of RQ2 (Section~\ref{sec:rq2-bug-finding-results}),
with a \empirical{30}-minute timeout.
We count frontend time as part of lowering.

\point{Results}
\tool\ is fast on most code
(Table~\ref{tab:performance-percentiles}).
At the median,
it analyzes an npm package in~\empirical{0.58}~s
and a PyPI module in~\empirical{0.45}~s,
and more than~\empirical{87\%} of the completed analyses
finish~\emph{within}~\empirical{10}~s.
Only~\empirical{22} PyPI modules
(\empirical{2\%})
and~\empirical{5} npm packages
(\empirical{0.7\%})
exceed the timeout
(Table~\ref{tab:performance-percentiles}
and Figure~\ref{fig:performance} exclude them).
These either statically link
entire third-party libraries
(e.g., SQLite in {\tt apsw}),
which slows down SVF,
or contain mutually recursive functions,
which slow down tag refinement.

Among the completed analyses,
lowering dominates,
with~\empirical{90\%} of the time in npm
and~\empirical{64\%} in PyPI.
In PyPI,
tag refinement takes~\empirical{32\%} of the time,
concentrated in a few modules:
five of them account for~\empirical{63\%} of it.
For example,
\texttt{zodbpickle.\_pickle} has
only~\empirical{17.9K} instructions,
yet tag refinement takes~\empirical{982}~s
of its~\empirical{\nnum{1116}}~s.
Its functions {\tt save}
and {\tt save\_dict} call each other,
and the analysis keeps
a copy of each statement for every route
through this recursion,
which multiplies the program points
that the fixpoint iteration visits.
Finally,
in both ecosystems,
the subtyping check is cheap:
it takes~\empirical{1.3\%} (npm) and~\empirical{1.4\%} (PyPI)
of the time,
and at most~\empirical{38}~s on a single module.

The analysis time grows with program size
(\cref{fig:performance}),
although the \texttt{zodbpickle} case shows
that tag refinement also depends on
the structure of the code.

\section{AI Use}
\label{sec:ai-use}

We used generative AI assistants in preparing this work.
Claude Code (Opus~5.5)
assisted with most of the implementation of~\tool.
We developed each feature with an explicit plan
and validated it through test-driven development.
To this end,
we assembled \empirical{427} micro-benchmarks,
each a small native extension with the expected output of~\tool.
Together with unit tests of the analysis components,
they form a suite of \empirical{831} tests.
Claude (Opus~5.5) also generated the scripts
that produce the paper's figures and tables from our evaluation data,
helped improve the writing,
and checked the manuscript for typographical and formatting errors.

Validating the~\empirical{\nnum{6920}} type errors
that~\tool\ reported required more work
than we could do by hand.
We employed an LLM agent
to construct a reproducer for each reported error
(Section~\ref{sec:rq2-bug-finding-results}).
However,
the agent may craft a reproducer
that triggers an error other than the reported one.
To guard against this,
we manually checked the reproducers of all
errors that we reported to developers
(\empirical{58} issues),
and confirmed that each triggers
the error reported by \tool.
The goal of this work is not to evaluate LLMs,
and a different model or different instructions
may perform better or worse.
\Cref{app:prompt} provides the prompt we used.

The research ideas,
the design of the approach,
and the interpretation of the results are our own.
We verified all generated output
and take full responsibility for the content of this paper.

\section{Related Work}
\label{sec:related}

\point{Native extensions and missing checks}
\Cref{sec:insights} explains why
the checkers of Furr and Foster~\cite{FF05,FF06,FF08}
do not apply to dynamic languages,
and why those of Brown et al.~\cite{brown-checker},
FlowJS~\cite{flowjs} and MOPSA~\cite{mopsa}
miss the errors we target.
\Cref{sec:rq1-effectiveness} evaluates \tool\
on the errors that the FlowJS study reports.
Other tools follow values across
the language boundary.
Lee et al.~\cite{LLR20} combine summaries of
the C code of JNI programs with an analysis of Java,
CHARON~\cite{charon} flags unsanitized flows
from scripts to native sinks
such as \code{strcpy} and \code{malloc},
and Gasket~\cite{gasket},
PyXray~\cite{pyxray} and AXA~\cite{RNHKM24}
identify the native functions a
script can call and the
values that cross the boundary.
None of them reasons about the tags of these values.

Other checkers target other rules of the binding API.
cpychecker~\cite{cpychecker} checks argument parsing
and reference counts,
Pungi~\cite{LT14} and PyRefcon~\cite{MYZYZ23}
check reference counts,
Li and Tan~\cite{LT09} check
the handling of JNI exceptions,
and Jinn~\cite{LWHGM10} checks JNI calls at run time.
CaVer~\cite{LSKL15},
TypeSan~\cite{HJHGKB16} and
HexType~\cite{JBCLP17}
detect type confusion between C++ classes.
A boundary type error is a type confusion between host tags,
which these tools do not track.

Some approaches find missing checks
without a given specification.
Engler et al.~\cite{ECHCC01} infer the expected checks
from common practice,
Crix~\cite{LPW19} from the checks in related contexts,
and Dillig et al.~\cite{DDA07}
from inconsistent uses of a value.
We instead take the tag that a use assumes
from the binding API specification.

\point{Occurrence typing, flow typing and semantic subtyping}
Tag refinement applies flow-sensitive type
refinement to native code.
It follows occurrence typing~\cite{THF08,THF10}
and flow typing~\cite{Pearce13},
which refine the type of a variable at each test.
Guha et al.~\cite{GSK11} reason about the tags
of JavaScript variables in single-language programs.
Our tags additionally include the classes
that native extensions register.

Our types denote sets of values,
and subtyping is set
inclusion~\cite{FCB08}.
CDuce~\cite{BCF03} and the SSTT library~\cite{LN26}
check this inclusion by testing
whether a type is empty,
and Dminor~\cite{BGHL12}
hands the check to an SMT solver.
We follow Dminor
and read a witness off the solver's model.



\section{Conclusion}
\label{sec:conclusion}

Native extensions make programs
in dynamic languages faster,
but they can break the type safety
of the host language.
We showed that native code contains
all the information needed to detect
such violations:
its tests establish which
tags a value may have,
and its uses state which tag they assume.
We implemented this idea in \tool,
which found thousands of confirmed errors
in npm and PyPI packages
and in the built-in extensions of CPython and Node.js.
Most errors arise because developers check
the arguments of a native function,
but not the values it obtains from the host,
or check them too broadly.
We have filed \empirical{58} issues
in \empirical{45} projects,
and \empirical{26} have been fixed.

\bibliographystyle{ACM-Reference-Format}
\bibliography{main}

\clearpage
\appendix
\crefalias{section}{appendix}
\section{Semantics of XLIR}
\label{app:semantics}

Figure~\ref{fig:sem} gives
the small-step semantics of XLIR
(\cref{sec:approach}).
A \emph{frame} $\Cfg{\env}{\sq}$ holds
the state of one function call: an environment
$\env$, which maps variables to values,
and the statements $\sq$ that remain to run.
A frame whose call waits for a callee
also records the variables
$\bar y$ that receive the results
of the callee, written
$\Cfg{\env}{\bar y\Leftarrow\sq}$.
A \emph{configuration} is a stack of
frames, written $\Cfg{\env}{\sq}\cdot K$,
with $\Cfg{\env}{\sq}$ on top.
Rules that show a single frame apply to the top frame and leave $K$
unchanged. A run of $P$ takes an argument vector
$\vec v=(v_0,\dots,v_{j-1})$,
none of which is $\emp$, and
$\Arg{i}$ reads $v_i$.
The run starts in $\Cfg{\{\}}{\sq_e}$,
where $\kw{fun}\ e()=\sq_e\in\bar F$.
It ends when the entry point returns,
or in the state $\wrong$ when a use is unsafe.

\begin{figure}[h]
\footnotesize
\setlength{\tabcolsep}{0pt}
\begin{tabular}{@{}l@{\hspace{2.5em}}l@{}}
$\inferrule*[lab=S-Arg]{\vec v=(v_0,\dots,v_{j-1})\in\Hs{Tag}^j \\ v=v_i\text{ if }i<j\text{, else }\tundef}
 {\Cfg{\env}{\Let{x}{\Arg{i}};\sq}\to\Cfg{\env[x\mapsto v]}{\sq}}$
&
$\inferrule*[lab=S-Hop]{\vdash v:\tau}
 {\Cfg{\env}{\Let{y}{\Host{\tau}{x}};\sq}\to\Cfg{\env[y\mapsto v]}{\sq}}$
\\[10pt]
$\inferrule*[lab=S-ChoiceT]{ }
 {\Cfg{\env}{\Ifs{\star}{\sq_1}{\sq_2};\sq}\to\Cfg{\env}{\sq_1\cat\sq}}$
&
$\inferrule*[lab=S-ChoiceF]{ }
 {\Cfg{\env}{\Ifs{\star}{\sq_1}{\sq_2};\sq}\to\Cfg{\env}{\sq_2\cat\sq}}$
\\[10pt]
$\inferrule*[lab=S-CheckT]{\vdash\env(x):\tau}
 {\Cfg{\env}{\Ifs{\ischk{\tau}{x}}{\sq_1}{\sq_2};\sq}\to\Cfg{\env}{\sq_1\cat\sq}}$
&
$\inferrule*[lab=S-CheckF]{\not\vdash\env(x):\tau}
 {\Cfg{\env}{\Ifs{\ischk{\tau}{x}}{\sq_1}{\sq_2};\sq}\to\Cfg{\env}{\sq_2\cat\sq}}$
\\[10pt]
$\inferrule*[lab=S-UseSafe]{\vdash\env(x):\tau}
 {\Cfg{\env}{\proj{\tau}{x};\sq}\to\Cfg{\env}{\sq}}$
&
$\inferrule*[lab=S-UseUnsafe]{\not\vdash\env(x):\tau}
 {\Cfg{\env}{\proj{\tau}{x};\sq}\cdot K\to\wrong}$
\\[10pt]
\multicolumn{2}{@{}l@{}}{%
$\inferrule*[lab=S-Call]{\kw{fun}\ f(\bar z)=\sq_f\in\bar F}
 {\Cfg{\env}{\Let{\bar y}{f(\bar x)};\sq}\cdot K\to\Cfg{[\bar z\mapsto\env(\bar x)]}{\sq_f}\cdot\Cfg{\env}{\bar y\Leftarrow\sq}\cdot K}$}
\\[10pt]
\multicolumn{2}{@{}l@{}}{%
$\inferrule*[lab=S-Ret]{ }
 {\Cfg{\env'}{\kw{return}\ \bar x;\sq'}\cdot\Cfg{\env}{\bar y\Leftarrow\sq}\cdot K\to\Cfg{\env[\bar y\mapsto\env'(\bar x)]}{\sq}\cdot K}$}
\end{tabular}
\caption{Small-step semantics of XLIR.}
\label{fig:sem}
\end{figure}

\section{Examples of Boundary Type Errors}
\label{app:examples}
We discuss three representative type errors
found by~\tool\
(\cref{sec:rq3-bug-characteristics}).
Figure~\ref{fig:example-bugs} contains
snippets of the buggy native code
and the corresponding reproducers.

\begin{figure}[t]
\renewcommand{\theFancyVerbLine}{\textcolor{black!40}{\sffamily\fontsize{6}{7}\selectfont\arabic{FancyVerbLine}}}%
\newcommand{\scriptlines}[1]{\renewcommand{\theFancyVerbLine}{\textcolor{black!40}{\sffamily\fontsize{6}{7}\selectfont
  \ifnum\value{FancyVerbLine}>#1 \else S\arabic{FancyVerbLine}\fi}}}%
\setminted{fontsize=\fontsize{6.6}{8.4}\selectfont, xleftmargin=11pt, numbersep=5pt, escapeinside=@@, formatcom=\codepalette}%
\renewcommand{\hlrad}{0pt}
\captionsetup[sub]{justification=centering,singlelinecheck=false,font=footnotesize}
\begin{minipage}[t]{0.315\textwidth}
\vspace{0pt}
\begin{minipage}[t]{\linewidth}
\scriptlines{6}%
\begin{minted}[linenos]{text}
@\replp@ ({ Document, Element } =
    require('libxmljs2'))
@\replp@ doc = new Document()
@\replp@ new Element(doc, 'SECRET_KEY')
@\replp@ e = new Element(doc, Symbol())
@\replp@ e.name()
@\replo{\textquotesingle SECRET\_KEY\textquotesingle}@
\end{minted}
\end{minipage}
\replsep
\begin{minipage}[t]{\linewidth}
\begin{minted}[linenos]{cpp}
NAN_METHOD(XmlElement::New) {
  XmlDocument *doc = ...;
  @\hb{badfill}@Nan::Utf8String s(info[1])@\he@;
  ...
  xmlNode *n = xmlNewDocNode(
      doc->xml_obj, NULL,
      *s, content);
  XmlElement *el =
      new XmlElement(n);
  el->Wrap(info.This());
}
\end{minted}
\end{minipage}
\end{minipage}
\hfill
\begin{minipage}[t]{0.315\textwidth}
\vspace{0pt}
\begin{minipage}[t]{\linewidth}
\scriptlines{5}%
\begin{minted}[linenos]{text}
@\replpy@ from google.protobuf \
@\replpyc@   import descriptor_pool
@\replpy@ class S(str): __len__ = None
@\replpy@ descriptor_pool.Default() \
@\replpyc@   .FindFileByName(S('a'))
@\replo{Segmentation fault}@
\end{minted}
\end{minipage}
\replsep
\begin{minipage}[t]{\linewidth}
\begin{minted}[linenos]{c}
PyObject *FindFileByName(
    Pool *self,
    PyObject *arg) {
  const char *name =
      PyUpb_VerifyStrData(arg);
  if (!name) return NULL;
  @\hb{badfill}@f = FindFileByNameWithSize(@\hc@
      self->symtab, name,
      PyObject_Size(arg))@\he@;
  ...
}
\end{minted}
\end{minipage}
\end{minipage}
\hfill
\begin{minipage}[t]{0.315\textwidth}
\vspace{0pt}
\begin{minipage}[t]{\linewidth}
\scriptlines{6}%
\begin{minted}[linenos]{text}
@\replp@ ({ sma } =
    require('tulind').indicators)
@\replp@ k = 0; xs = []
@\replp@ xs.__defineGetter__(0,
    () => k++ ? 42 : [1,2,3])
@\replp@ sma.indicator(xs, [3])
@\replo{Segmentation fault}@
\end{minted}
\end{minipage}
\replsep
\begin{minipage}[t]{\linewidth}
\begin{minted}[linenos]{cpp}
for (i = 0; i < len; ++i) {
  auto v = Nan::Get(inputs, i)
      .ToLocalChecked();
  if (@\hb{weakfill}@!v->IsArray()@\he@) return;
}
for (i = 0; i < len; ++i) {
  @\hb{badfill}@auto a = Local<Array>::Cast(@\hc@
      Nan::Get(inputs, i)
          .ToLocalChecked())@\he@;
  offset = a->Length() - size;
}
\end{minted}
\end{minipage}
\end{minipage}

\vspace{4pt}
\begin{minipage}[t]{0.315\textwidth}
\subcaption{\texttt{libxmljs2} 0.37.0}
\label{fig:bug-libxmljs2}
\end{minipage}
\hfill
\begin{minipage}[t]{0.315\textwidth}
\subcaption{\texttt{protobuf} 7.36.2}
\label{fig:bug-protobuf}
\end{minipage}
\hfill
\begin{minipage}[t]{0.315\textwidth}
\subcaption{\texttt{tulind} 0.8.20}
\label{fig:bug-tulind}
\end{minipage}
\caption{Three type errors found by~\tool,
each with an excerpt of the native code
and a JavaScript or Python reproducer.}
\label{fig:example-bugs}
\end{figure}

\point{\cref{fig:bug-libxmljs2}
(\href{https://www.npmjs.com/package/libxmljs2}{\texttt{libxmljs2}})}
The native function \texttt{XmlElement::New} implements
the JavaScript constructor
\texttt{new Element(doc, name)}
of the \texttt{libxmljs2} package.
It converts its second argument,
the name of the element,
to a UTF-8 string (line~3)
and uses it to create an XML node (lines~5--7),
which it wraps into the returned object (lines~8--10).
The code neither checks the type of the argument
nor whether the conversion succeeds
(\textit{no check}).
When the name is a symbol
(see the reproducer),
the conversion fails
and leaves its stack buffer unwritten.
In this case,
the node is named with whatever
an earlier call left in that buffer.
As the reproducer shows,
\texttt{name()} then returns this stale
string to JavaScript,
which may contain sensitive data
(\textit{information disclosure}).

\point{\cref{fig:bug-protobuf}
(\href{https://pypi.org/project/protobuf/}{\texttt{protobuf}})}
The native function \texttt{FindFileByName} implements
the Python method \texttt{DescriptorPool.FindFileByName(name)}
of the \texttt{protobuf} package,
which looks up a \texttt{.proto} file by its name.
It checks that the name is a \texttt{str} or \texttt{bytes} object
and gets its character data (lines~4--6).
It then looks up the name in a hash table,
passing its data and its length (lines~7--9).
However,
the length comes from the host operation~\texttt{PyObject\_Size} (line~9),
which calls the~\texttt{\_\_len\_\_} method of the name.
Since the check \texttt{PyUpb\_VerifyStrData} (line~5)
permits subclasses of~\texttt{str},
\texttt{\_\_len\_\_} can be any Python code.
When it fails,
\texttt{PyObject\_Size} returns~$-1$,
but the code never checks this result
(\textit{no check}).
The lookup then reads the name
with a length of \texttt{(size\_t)-1},
far past its end,
and the process crashes with a segmentation fault
(\textit{hard crash}).

\point{\cref{fig:bug-tulind}
(\href{https://www.npmjs.com/package/tulind}{\texttt{tulind}})}
The functions in the {\tt indicators} property of
the \texttt{tulind} package
take an array of input arrays.
A first loop reads each element
of the input container (lines~2--3)
and rejects it unless it is an array (line~4).
A second loop reads each element again (lines~7--9),
casts it to an array without a check,
and calls \texttt{Length()} on it (line~10).
Therefore,
the check holds only for the first read
(\textit{incomplete check}).
This is because
reading an element may invoke a JavaScript getter,
which can return a different value each time.
In the reproducer,
the getter returns an array on the first read
and a number on the second,
so line~10 calls \texttt{Length()} on a number,
and the process crashes with a segmentation fault.

\section{False Negatives and False Positives in~\tool}
\label{app:fn-fp}
\point{False negatives}
\tool\ misses type errors
that lie in code it never analyzes.
Its analysis follows the control flow
of a native entry function,
so it does not examine
(1) callbacks that run asynchronously,
after the entry function has returned,
or (2) functions invoked by C libraries
whose code is not available to the analysis.
Also,
some boundary values reach type-unsafe operations
through binding API functions
that~\tool{} does not model.

\Cref{fig:false-negative} illustrates
a false negative in {\tt libpq}.
The native entry {\tt Connect} casts its second argument
to a function without checking its type (lines~2--3),
stores it in an asynchronous worker (lines~4--9),
and schedules the worker (line~10).
The entry then returns.
Later,
the event loop runs {\tt HandleOKCallback},
which calls the stored value (line~15)
and crashes the process
when the argument is not a function.
Because the event loop,
rather than the entry function,
invokes {\tt HandleOKCallback} (lines~13--16),
\tool\ never analyzes it
and misses the bug.
Detecting such bugs requires modeling
how each binding framework schedules deferred work
(e.g., {\tt Nan::AsyncQueueWorker}).
We leave it as future work.

\begin{figure}[t]
\renewcommand{\theFancyVerbLine}{\textcolor{black!40}{\sffamily\fontsize{6}{7}\selectfont\arabic{FancyVerbLine}}}%
\setminted{escapeinside=@@, formatcom=\codepalette}%
\renewcommand{\hlrad}{0pt}
\begin{minipage}[t]{0.49\textwidth}
\begin{minted}[linenos]{cpp}
NAN_METHOD(Connection::Connect) {
  v8::Local<v8::Function> callback =
      info[1].As<v8::Function>();
  Nan::Callback* nanCallback =
      new Nan::Callback(callback);
  ConnectAsyncWorker* worker =
      new ConnectAsyncWorker(
          info[0].As<v8::String>(),
          self, nanCallback);
  Nan::AsyncQueueWorker(worker);
}

void ConnectAsyncWorker::HandleOKCallback() {
  // crashes if callback is not a function.
  @\hb{badfill}@callback->Call(0, NULL, async_resource)@\he@;
}
\end{minted}
\end{minipage}
\hfill
\begin{minipage}[t]{0.49\textwidth}
\begin{minted}[linenos]{c}
void Object_beginTypeContext(JSOBJ obj,
    JSONTypeContext *tc) {
  pc = createTypeContext(); tc->prv = pc;
  ...
  } else if (@\hb{weakfill}@PyBytes_Check(obj)@\he@) {
    pc->PyTypeToUTF8 = PyBytesToUTF8;
  } ...
}
char *PyBytesToUTF8(JSOBJ _obj, ...) {
  PyObject *obj = (PyObject *)_obj;
  // obj is always bytes here (line 5)
  *_outLen = @\hb{badfill}@PyBytes_GET_SIZE(obj)@\he@;
  return @\hb{badfill}@PyBytes_AS_STRING(obj)@\he@;
}
char *Object_getStringValue(JSOBJ obj,
    JSONTypeContext *tc, ...) {
  return GET_TC(tc)->PyTypeToUTF8(...);
}
\end{minted}
\end{minipage}

\begin{minipage}[t]{0.49\textwidth}
\subcaption{
False negative in \texttt{libpq} 1.11.0.}
\label{fig:false-negative}
\end{minipage}
\hfill
\begin{minipage}[t]{0.49\textwidth}
\subcaption{
False positive in \texttt{pandas} 1.4.1.
}
\label{fig:false-positive}
\end{minipage}
\caption{Examples of false negatives and false positives
exhibited by~\tool.}
\label{fig:fn-fp}
\end{figure}

\point{False positives}
Most false positives are reports on paths
that cannot execute with wrongly-typed inputs,
and the analysis cannot see why
a use is type-safe.
This includes
paths guarded by native conditions that never hold,
or uses that are reached only after
another argument has already caused a crash.
Another source of false positives
is that~\tool\ loses information
where control-flow paths merge.
For example,
a variable may hold a JavaScript value
that has been type-checked on one branch,
and a default value created in
native code on the other.
Later uses of the variable are
safe on both paths,
but for different reasons.
In the first case,
the value has been type-checked,
and in the second,
native code has created it with the right type.
However,
after the branches merge,
\tool\ keeps only what holds on~\emph{every} path,
so it forgets that the value is safe
and reports its later uses.
Similarly,
when a C/C++ function is reached through call chains
that share their $k$ most recent call sites,
\tool\ merges what it learns from these chains
(\cref{app:calls}),
so a value checked on only some of them
is treated as unchecked on all of them.

\Cref{fig:false-positive} shows
a false positive in {\tt pandas}.
The function {\tt PyBytesToUTF8}
reads its argument as a {\tt bytes} object
(lines~12--13).
The function is stored in a field of
a heap-allocated struct
only on the branch where
{\tt PyBytes\_Check(obj)} holds
(lines~3--6),
and {\tt Object\_getStringValue} later calls it
through that field,
which it reaches by reloading {\tt tc->prv}
through {\tt GET\_TC(tc)}
(line~17).
Hence,
the argument is always a {\tt bytes} object
when the callback runs.
\tool\ does not track type information
through values stored in heap memory.
Therefore,
it cannot tell that the value
was checked before the pointer was saved,
and it reports the reads of the value as type-unsafe.

\section{Object Writes in XLIR}
\label{app:writes}
Native code can also write to host values and create new ones
(Table~\ref{tab:writes}).
XLIR has no statement for either.
Neither changes the tag of an existing value,
and a boundary type error depends only on the tag
of the value that reaches a use.

\begin{table}[h]
\vspace{-0.5\baselineskip}
\caption{What native code can do to host values, and what it does to tags.}
\label{tab:writes}
\footnotesize
\renewcommand{\code}[1]{{\scriptsize\texttt{#1}}}%
\begin{tabular}{@{}llll@{}}
\toprule
\textbf{Operation} & \textbf{V8, N-API} & \textbf{CPython} & \textbf{Tag}\\
\midrule
Write a member & \code{obj->Set(ctx, k, v)}, \code{napi\_set\_property} & \code{PyObject\_SetAttr}, \code{PyList\_SetItem} & unchanged\\
Delete a member & \code{obj->Delete(ctx, k)} & \code{PyObject\_DelAttr} & unchanged\\
Change the prototype & \code{obj->SetPrototype(ctx, p)} & --- & unchanged\\
Create a value & \code{Array::New}, \code{napi\_create\_array} & \code{PyLong\_FromLong}, \code{PyList\_New} & a new value\\
\bottomrule
\end{tabular}
\vspace{-0.5\baselineskip}
\end{table}

\section{Function Calls in Tag Refinement}
\label{app:calls}
The analysis is context-sensitive, following the call-strings
approach~\cite{sharir1981}. A context is the $k$ most recent call sites that
lead to a function, and we analyze the function once for each context. A
call
passes the types of its arguments to the parameters of the callee, and gives
$\bar y$ the types $\mathit{ret}_f$ of the variables that the callee returns
in that context, joined over its return statements (\cref{fig:taginf-rules}). Calls that
reach a function through the same $k$ call sites share its analysis, so
their argument types are joined. For example, consider
$\kw{fun}\ \mathit{id}(z)=\kw{return}\ z$, called as
$\Let{a}{\mathit{id}(x)}$ where $\Gamma(x)=\tstr$, and as
$\Let{b}{\mathit{id}(y)}$ where $\Gamma(y)=\Any$. If $\mathit{id}$ were
analyzed once, $z$ and hence $a$ would have type $\tstr\tor\Any$, which is
equivalent to $\Any$, and a later $\proj{\tstr}{a}$ would be reported. With one analysis
per context, $a$ has type $\tstr$.

\section{Instructions Given to the Validation Agent of RQ1 and RQ2}
\label{app:prompt}

We ran the validation agent of RQ1 and RQ2
in Claude Code with Claude Opus 5.5 at high reasoning effort
(\cref{sec:rq1-effectiveness,sec:rq2-bug-finding-results}).
A run covers one native extension
and up to five of the type errors that~\tool\ reports in it,
and lasts at most 30 minutes.
Each run receives the instructions,
the reports to validate,
the source code of the extension
and its compiled module.
In RQ1,
the agent validates only the reports beyond the known type errors,
and it also receives the reproducers of the known errors
as examples of how to call the module.

Figure~\ref{fig:prompt} gives the instructions in simplified form.
We shortened the wording,
and \code{[...]} marks text that we cut.
The cut text covers the commands that the agent may run,
the names and the layout of the files,
the format of the answer,
and the detailed definitions of the terms, with their examples.
In the instructions,
the requirement of an operation is the tag that a use assumes.
A reproducer calls the native entry points directly,
in line with the most-general client assumption
(\cref{sec:bg-apis}).

The agent gives one verdict per report.
We count a report as confirmed
when the agent provides a reproducer for it.
For a latent finding,
the reproducer triggers an earlier error on the same path,
which hides the reported one.
An unreachable finding has no reproducer.
We count it as unconfirmed,
together with the verdicts \code{false\_positive} and \code{unresolved}.
A script runs every reproducer and its control again,
and rejects a finding whose reproducer does not show the stated symptom
or whose control does not exit cleanly.
The control differs from the reproducer
only in the value at the reported use,
so a symptom that appears only in the reproducer
is caused by that value.

\begin{figure}[tp]
\begin{promptbox*}
\prompthead{Task}
You work on one native extension module, in your working directory. Xtypecheck reported type errors
in it. Each report names an operation in the native code and a requirement that the operation
relies on about a boundary value: that the value has a type, that an operation on it did not fail,
or that no exception is pending. For each report, look for the evidence of a bug, or for the check
that protects the operation.~\elided
\prompthead{Inputs}
The reports are at path: <reports>. Each has a location, the enclosing function, the requirement
(what the native code relies on without checking), a witness, its entry points, and
some of the call chains that reach it.

The source code is at path: <source>.

The compiled module is at path: <module>.~\elided
\prompthead{When a report is genuine}
A report is genuine exactly when (1) the requirement false makes the operation fail \elided; and
(2) on some path through the code to the
operation, whether or not an input can drive execution along it, no check establishes the
requirement before the first operation that relies on it.~\elided
\prompthead{Verdicts}
\promptitem{finding:}{You have the evidence of a bug, in one of three forms.}
\promptsubitem{reproduced:}{A program P, calling only the module's native entries with anything the host
  language allows, reaches the operation with the requirement false and shows a
  symptom that its control P' does not. P' is P with only the value replaced by one
  that meets the requirement.~\elided}
\promptsubitem{latent:}{Every value that makes the requirement false fails earlier on the same path, at an
  operation that is itself a bug, and no check of the requirement lies between that
  operation and this one. The reproducer is the earlier failure's.~\elided}
\promptsubitem{unreachable:}{No input reaches the operation, because no entry reaches its function or the path
  needs native state that no input can create. The code is the evidence.}
\promptitem{false\_positive:}{For every path from every entry, you can name the check that establishes the
  requirement in time, or why the requirement false cannot make the operation fail.}
\promptitem{unresolved:}{You have neither.~\elided}
\prompthead{Symptoms}
A hard crash. An unintended side effect: native code changes state after a failure it did not test.
Incorrect output: a value that does not come from the input. An exception that is left pending or
lost. A failure that native code reports correctly to the host is not a symptom.~\elided
\prompthead{Finding evidence}
Before any false\_positive or unresolved, try each of the following that applies:
\promptbullet{the value at the top level, and nested in each container the entry accepts \elided;}
\promptbullet{each entry that reaches the operation, not only the first one;}
\promptbullet{the call repeated, and other calls on the same object or module after it;}
\promptbullet{a subclass of the expected type, and an object that forges the type check \elided;}
\promptbullet{methods that raise, and methods that return a value of the wrong type.}
\elided{} A read of another type's memory or past an object often survives without a crash. Where the
benchmark has an AddressSanitizer build, run the reproducer under it.~\elided

The witness is a starting point for building an input. It is not evidence of an error. A program that
shows no symptom proves nothing either way.
\prompthead{Reproducers}
\elided{} A reproducer loads the module directly, never through the package's own code. \elided{} Run with
the argument control, it makes the same calls with a valid value at the failing operation and exits
0 with no symptom. \elided{} One failing operation per process, so a symptom belongs to its finding.
Do not build, compile or install anything, and do not use the network.~\elided
\end{promptbox*}
\caption{Instructions of the validation agent.}
\label{fig:prompt}
\end{figure}

\section{Calling Conventions of Entry Points}
\label{app:conventions}

\newcommand{\convsetup}{%
\scriptsize
\renewcommand{\arraystretch}{0.9}%
\setlength{\tabcolsep}{3pt}%
\renewcommand{\code}[1]{{\scriptsize\texttt{##1}}}%
\newcommand{\sig}[2]{{\scriptsize\ttfamily\begin{tabular}[t]{@{}>{\scriptsize\ttfamily}l@{}}##1(\\\hspace*{1em}\begin{tabular}[t]{@{}>{\scriptsize\ttfamily}l@{}}##2)\end{tabular}\end{tabular}}}%
\newcommand{\bv}[1]{\textbf{##1}}%
\newcommand{\apitag}[1]{\tikz[baseline=(t.base)]\node[fill=black!8,rounded corners=1pt,inner xsep=1.5pt,inner ysep=1pt](t){\scriptsize\ttfamily\strut##1};}%
\newcommand{\kind}[2]{\begin{tabular}[t]{@{}l@{}}##1\\\apitag{##2}\end{tabular}}%
\newcommand{\kindm}[2]{\begin{tabular}[t]{@{}l@{}}##1\\##2\end{tabular}}%
\newcommand{\softrule}{\arrayrulecolor{black!25}\midrule\arrayrulecolor{black}}%
}

Table~\ref{tab:entries} (\cref{sec:entry-point}) shows six of the calling conventions
that the frontend of~\tool\ recognizes.
This section lists all of them.
Table~\ref{tab:conv-js} covers the three binding APIs of Node.js,
Table~\ref{tab:conv-py} the methods and attributes of CPython,
and Table~\ref{tab:conv-slots} the type slots of CPython.
The tables use the notation of Table~\ref{tab:entries}.
Each row is a rule,
the grey label under a kind is its registration point,
bold parameters carry boundary values,
and $\bar{x}$ stands for $x_0, x_1, \dots$.

In NAN and node-addon-api,
an entry point reads its receiver and its arguments
from its parameter \code{info},
and a setter receives the assigned value as a second parameter.
In N-API,
an entry point first calls
\code{napi\_get\_cb\_info(env, info, \&argc, argv, ...)},
which copies its arguments into the array \code{argv}.

In CPython,
the flags of a method (\code{ml\_flags}) select its convention.
A \code{METH\_VARARGS} method receives its arguments in the tuple \code{p1}.
It reads argument~$k$ with \code{PyTuple\_GET\_ITEM},
or with a parser such as \code{PyArg\_ParseTuple},
which stores argument~$k$ through the output pointer of its $k$-th format unit.
\code{PyArg\_ParseTupleAndKeywords} stores keyword arguments in the same way.
The flags \code{METH\_CLASS} and \code{METH\_STATIC}
combine with every row of Table~\ref{tab:conv-py}.
With them, \code{p0} is the class or null and carries no boundary value.
The same holds for a function of a module, where \code{p0} is the module.
Table~\ref{tab:conv-slots} groups the type slots
that share their parameters and their boundary values.

\begin{table}[tp]
\caption{Calling conventions of entry points in the binding APIs of Node.js.}
\label{tab:conv-js}
\vspace{-5pt}
\convsetup
\begin{tabularx}{\linewidth}{@{}llll>{\raggedright\arraybackslash}X@{}}
\toprule
\textbf{Binding API} & \textbf{Kind} & \textbf{Host call} & \textbf{C/C++ signature} & \textbf{Boundary values}\\
\midrule
N-API
  & \kindm{function, method, constructor}{\apitag{napi\_create\_function}\\\apitag{napi\_define\_properties}\\\apitag{napi\_define\_class}}
  & $f(\bar{x})$
  & \sig{napi\_value f}{napi\_env env,\\napi\_callback\_info \bv{info}}
  & \mbox{$x_k$ = \code{argv[k]}}\\
\softrule
NAN
  & \kind{function}{SetMethod} & $f(\bar{x})$
  & \sig{void f}{const Nan::FunctionCallbackInfo\\<Value>\& \bv{info}}
  & \mbox{$x_k$ = \code{\bv{info}[k]}}\\
  & \kind{method}{SetPrototypeMethod} & $o.f(\bar{x})$
  & \sig{void f}{const Nan::FunctionCallbackInfo\\<Value>\& \bv{info}}
  & \mbox{$o$ = \code{\bv{info}.This()}}, \mbox{$x_k$ = \code{\bv{info}[k]}}\\
  & \kind{constructor}{New<FunctionTemplate>} & $\mathtt{new}\ C(\bar{x})$
  & \sig{void f}{const Nan::FunctionCallbackInfo\\<Value>\& \bv{info}}
  & \mbox{$x_k$ = \code{\bv{info}[k]}}\\
  & \kind{getter}{SetAccessor} & $o.p$
  & \sig{void g}{Local<String> p,\\const Nan::PropertyCallbackInfo\\<Value>\& \bv{info}}
  & \mbox{$o$ = \code{\bv{info}.This()}}\\
  & \kind{setter}{SetAccessor} & $o.p = v$
  & \sig{void s}{Local<String> p,\\Local<Value> \bv{v},\\const Nan::PropertyCallbackInfo\\<void>\& \bv{info}}
  & \mbox{$o$ = \code{\bv{info}.This()}}, \mbox{$v$ = \code{\bv{v}}}\\
  & \kindm{property handler}{\apitag{SetNamedPropertyHandler}\\\apitag{SetIndexedPropertyHandler}}
  & $o[k]$, $o[k] = v$
  & as for getter and setter
  & \mbox{$o$ = \code{\bv{info}.This()}}, \mbox{$v$ = \code{\bv{v}}}\\
\softrule
\multirow{2}{*}{\begin{tabular}{@{}l@{}}node-\\addon-api\end{tabular}}
  & \kind{function}{Function::New} & $f(\bar{x})$
  & \sig{Napi::Value f}{const Napi::CallbackInfo\& \bv{info}}
  & \mbox{$x_k$ = \code{\bv{info}[k]}}\\
  & \kind{method}{InstanceMethod} & $o.f(\bar{x})$
  & \sig{Napi::Value T::f}{const Napi::CallbackInfo\& \bv{info}}
  & \mbox{$o$ = \code{\bv{info}.This()}}, \mbox{$x_k$ = \code{\bv{info}[k]}}\\
  & \kind{static method}{StaticMethod} & $C.f(\bar{x})$
  & \sig{Napi::Value T::f}{const Napi::CallbackInfo\& \bv{info}}
  & \mbox{$x_k$ = \code{\bv{info}[k]}}\\
  & \kind{constructor}{DefineClass} & $\mathtt{new}\ C(\bar{x})$
  & \sig{T::T}{const Napi::CallbackInfo\& \bv{info}}
  & \mbox{$x_k$ = \code{\bv{info}[k]}}\\
  & \kind{getter}{InstanceAccessor} & $o.p$
  & \sig{Napi::Value T::g}{const Napi::CallbackInfo\& \bv{info}}
  & \mbox{$o$ = \code{\bv{info}.This()}}\\
  & \kind{setter}{InstanceAccessor} & $o.p = v$
  & \sig{void T::s}{const Napi::CallbackInfo\& \bv{info},\\const Napi::Value\& \bv{v}}
  & \mbox{$o$ = \code{\bv{info}.This()}}, \mbox{$v$ = \code{\bv{v}}}\\
  & \kind{static setter}{StaticAccessor} & $C.p = v$
  & \sig{void T::s}{const Napi::CallbackInfo\& \bv{info},\\const Napi::Value\& \bv{v}}
  & \mbox{$v$ = \code{\bv{v}}}\\
\bottomrule
\end{tabularx}
\end{table}

\begin{table}[tp]
\caption{Calling conventions of methods and attributes in CPython.}
\label{tab:conv-py}
\vspace{-5pt}
\convsetup
\begin{tabularx}{\linewidth}{@{}lll>{\raggedright\arraybackslash}X@{}}
\toprule
\textbf{Kind} & \textbf{Host call} & \textbf{C signature} & \textbf{Boundary values}\\
\midrule
\kind{method}{METH\_NOARGS} & $o.f()$
  & \sig{PyObject *f}{PyObject *\bv{p0},\\PyObject *p1}
  & \mbox{$o$ = \code{\bv{p0}}}\\
\kind{method}{METH\_O} & $o.f(x)$
  & \sig{PyObject *f}{PyObject *\bv{p0},\\PyObject *\bv{p1}}
  & \mbox{$o$ = \code{\bv{p0}}}, \mbox{$x$ = \code{\bv{p1}}}\\
\kind{method}{METH\_VARARGS} & $o.f(\bar{x})$
  & \sig{PyObject *f}{PyObject *\bv{p0},\\PyObject *\bv{p1}}
  & \mbox{$o$ = \code{\bv{p0}}}, \mbox{$x_k$ = \code{PyTuple\_GET\_ITEM(\bv{p1},k)}}\\
\kind{method}{METH\_VARARGS|METH\_KEYWORDS} & $o.f(\bar{x}, n = v)$
  & \sig{PyObject *f}{PyObject *\bv{p0},\\PyObject *\bv{p1},\\PyObject *p2}
  & \mbox{$o$ = \code{\bv{p0}}}, \mbox{$x_k$ = \code{PyTuple\_GET\_ITEM(\bv{p1},k)}}\\
\kind{method}{METH\_FASTCALL} & $o.f(\bar{x})$
  & \sig{PyObject *f}{PyObject *\bv{p0},\\PyObject *const *\bv{p1},\\Py\_ssize\_t n}
  & \mbox{$o$ = \code{\bv{p0}}}, \mbox{$x_k$ = \code{\bv{p1}[k]}}\\
\kind{method}{METH\_FASTCALL|METH\_KEYWORDS} & $o.f(\bar{x}, n = v)$
  & \sig{PyObject *f}{PyObject *\bv{p0},\\PyObject *const *\bv{p1},\\Py\_ssize\_t n,\\PyObject *kw}
  & \mbox{$o$ = \code{\bv{p0}}}, \mbox{$x_k$ = \code{\bv{p1}[k]}}\\
\kindm{method}{\apitag{METH\_METHOD|METH\_FASTCALL|}\\\apitag{METH\_KEYWORDS}} & $o.f(\bar{x}, n = v)$
  & \sig{PyObject *f}{PyObject *\bv{p0},\\PyTypeObject *cls,\\PyObject *const *\bv{p2},\\Py\_ssize\_t n,\\PyObject *kw}
  & \mbox{$o$ = \code{\bv{p0}}}, \mbox{$x_k$ = \code{\bv{p2}[k]}}\\
\softrule
\kind{getter}{PyGetSetDef.get} & $o.p$
  & \sig{PyObject *g}{PyObject *\bv{p0},\\void *c}
  & \mbox{$o$ = \code{\bv{p0}}}\\
\kind{setter}{PyGetSetDef.set} & $o.p = v$
  & \sig{int s}{PyObject *\bv{p0},\\PyObject *\bv{p1},\\void *c}
  & \mbox{$o$ = \code{\bv{p0}}}, \mbox{$v$ = \code{\bv{p1}}}\\
\bottomrule
\end{tabularx}
\end{table}

\begin{table}[tp]
\caption{Calling conventions of type slots in CPython.
\code{p0}, \code{p1} and \code{p2} have the type \mbox{\code{PyObject *}}.}
\label{tab:conv-slots}
\vspace{-5pt}
\convsetup
\begin{tabularx}{\linewidth}{@{}>{\raggedright\arraybackslash\scriptsize\ttfamily}p{126pt}>{\raggedright\arraybackslash}p{58pt}l>{\raggedright\arraybackslash}X@{}}
\toprule
\textnormal{\textbf{Slots}} & \textbf{Host call} & \textbf{Parameters} & \textbf{Boundary values}\\
\midrule
tp\_repr, tp\_str, tp\_hash, tp\_iter, tp\_iternext, tp\_finalize,
am\_await, am\_aiter, am\_anext,
nb\_negative, nb\_positive, nb\_absolute, nb\_invert,
nb\_int, nb\_float, nb\_index, nb\_bool,
mp\_length, sq\_length
  & $\mathtt{repr}(o)$, $\mathtt{len}(o)$, $-o$, \ldots
  & \code{(\bv{p0})}
  & \mbox{$o$ = \code{\bv{p0}}}\\
\softrule
tp\_getattro, mp\_subscript, sq\_concat, sq\_contains, sq\_inplace\_concat,
nb\_inplace\_add, nb\_inplace\_subtract, nb\_inplace\_multiply,
nb\_inplace\_remainder, nb\_inplace\_lshift, nb\_inplace\_rshift,
nb\_inplace\_and, nb\_inplace\_xor, nb\_inplace\_or,
nb\_inplace\_floor\_divide, nb\_inplace\_true\_divide,
nb\_inplace\_matrix\_multiply
  & $o.x$, $o[x]$, $o + x$, $x\ \mathtt{in}\ o$, $o \mathrel{+}= x$, \ldots
  & \code{(\bv{p0}, \bv{p1})}
  & \mbox{$o$ = \code{\bv{p0}}}, \mbox{$x$ = \code{\bv{p1}}}\\
\softrule
tp\_setattro, mp\_ass\_subscript, tp\_descr\_set, nb\_inplace\_power
  & $o.x = v$, $o[x] = v$, \ldots
  & \code{(\bv{p0}, \bv{p1}, \bv{p2})}
  & \mbox{$o$ = \code{\bv{p0}}}, \mbox{$x$ = \code{\bv{p1}}}, \mbox{$v$ = \code{\bv{p2}}}\\
\softrule
tp\_richcompare
  & $o < x$, $o == x$, \ldots
  & \code{(\bv{p0}, \bv{p1}, int op)}
  & \mbox{$o$ = \code{\bv{p0}}}, \mbox{$x$ = \code{\bv{p1}}}\\
\softrule
tp\_descr\_get
  & $o.\mathtt{\_\_get\_\_}(x, t)$
  & \code{(\bv{p0}, \bv{p1}, p2)}
  & \mbox{$o$ = \code{\bv{p0}}}, \mbox{$x$ = \code{\bv{p1}}}\\
\softrule
nb\_add, nb\_subtract, nb\_multiply, nb\_remainder, nb\_divmod,
nb\_lshift, nb\_rshift, nb\_and, nb\_xor, nb\_or,
nb\_floor\_divide, nb\_true\_divide, nb\_matrix\_multiply
  & $a + b$, $a - b$, \ldots
  & \code{(\bv{p0}, \bv{p1})}
  & \mbox{$a$ = \code{\bv{p0}}}, \mbox{$b$ = \code{\bv{p1}}}\\
\softrule
nb\_power
  & $\mathtt{pow}(a, b, m)$
  & \code{(\bv{p0}, \bv{p1}, \bv{p2})}
  & \mbox{$a$ = \code{\bv{p0}}}, \mbox{$b$ = \code{\bv{p1}}}, \mbox{$m$ = \code{\bv{p2}}}\\
\softrule
tp\_call, tp\_init
  & $o(\bar{x})$, $C(\bar{x})$
  & \code{(\bv{p0}, \bv{p1}, p2)}
  & \mbox{$o$ = \code{\bv{p0}}}, \mbox{$x_k$ = \code{PyTuple\_GET\_ITEM(\bv{p1},k)}}\\
\softrule
tp\_new
  & $C(\bar{x})$
  & \code{(PyTypeObject *t, \bv{p1}, p2)}
  & \mbox{$x_k$ = \code{PyTuple\_GET\_ITEM(\bv{p1},k)}}\\
\softrule
sq\_repeat, sq\_inplace\_repeat, sq\_item
  & $o * n$, $o \mathrel{*}= n$, $o[i]$
  & \code{(\bv{p0}, Py\_ssize\_t i)}
  & \mbox{$o$ = \code{\bv{p0}}}\\
\softrule
sq\_ass\_item
  & $o[i] = v$
  & \code{(\bv{p0}, Py\_ssize\_t i, \bv{p2})}
  & \mbox{$o$ = \code{\bv{p0}}}, \mbox{$v$ = \code{\bv{p2}}}\\
\softrule
bf\_getbuffer, bf\_releasebuffer
  & $\mathtt{memoryview}(o)$
  & \code{(\bv{p0}, Py\_buffer *b, ...)}
  & \mbox{$o$ = \code{\bv{p0}}}\\
\bottomrule
\end{tabularx}
\end{table}

\end{document}

%% file: includes.tex
\usepackage{balance}
\usepackage{tabularx}
\usepackage{url}
\usepackage{graphicx}
\usepackage{listings}
\usepackage{enumitem}
\usepackage{amsmath}
\usepackage{pifont}
\usepackage{bigstrut}
\usepackage{mathpartir}
\usepackage[altpo]{backnaur}
\usepackage{nccmath}
\usepackage{sepnum}
\usepackage{caption}
\usepackage{threeparttable}
\usepackage[linesnumbered,ruled,noend]{algorithm2e}
\usepackage{natbib}
\usepackage{microtype}
\usepackage{wrapfig}
\usepackage{subcaption}
\usepackage{multirow}
\usepackage{multicol}
\usepackage{cleveref}
\usepackage[framemethod=TikZ]{mdframed}
\usepackage{colortbl}
\usepackage{arydshln}
\usepackage{parcolumns} 

\usepackage[frozencache]{minted2}
\usepackage[most]{tcolorbox} 
\usepackage{pifont}
\usepackage{comment}

\SetKw{Break}{break}
\SetKw{Continue}{continue}
\newcounter{example}

\newcounter{boxedd}

\definecolor{eggshell}{rgb}{0.94, 0.92, 0.84}
\definecolor{floralwhite}{rgb}{1.0, 0.98, 0.94}
\definecolor{blush}{rgb}{1, 0.5, 0.5}

\mdfdefinestyle{example}{%
	frametitle=\mbox{},
	frametitlerulewidth=0.00pt,
	frametitlebelowskip=0.00pt,
	frametitlerule=false,
    frametitlefont={\footnotesize},
    backgroundcolor=floralwhite!1,
	innerleftmargin=0.5pt,
	innerrightmargin=3pt,
    linecolor=eggshell,
    topline=false,
    bottomline=true,
    leftline=false,rightline=true,
    outerlinewidth=0.5pt,
    roundcorner=2.5mm,
    skipabove=0.0pt,
    skipbelow=0.0pt,
}
\newmdenv[%
    style=example,
    settings={\global\refstepcounter{example}},
    frametitlefont={\bfseries\small Example bug:\quad},
]{example}

\SetCommentSty{mycommfont}
\PassOptionsToPackage{hyphens}{url}\usepackage{hyperref}

\setcopyright{rightsretained}
\setlist[itemize]{noitemsep, leftmargin=1em}
\setlist[enumerate]{noitemsep, leftmargin=1em}
\newcommand{\nnum}[1]{\sepnum{.}{,}{}{#1}}

\setlist[itemize]{noitemsep, leftmargin=1em}
\setlist[enumerate]{noitemsep, leftmargin=1em}

\definecolor{dkred}{RGB}{87,10,10}
\definecolor{burntorange}{rgb}{0.8, 0.33, 0.0}
\definecolor{OrangeRed}{rgb}{1.0, 0.27, 0.0}
\definecolor{ForestGreen}{rgb}{0.0, 0.27, 0.13}

\definecolor{groovyblue}{HTML}{0000A0}
\definecolor{groovygreen}{HTML}{008000}
\definecolor{darkgray}{rgb}{.4,.4,.4}

\lstdefinelanguage{diff}{
  basicstyle=\ttfamily\small,
  morecomment=[f][\color{diffstart}]{@@},
  morecomment=[f][\bfseries\color{groovygreen}]{+\ },
  morecomment=[f][\bfseries\color{dkred}]{-\ },                    
}

\lstdefinelanguage{puppet}{
  morecomment = [l]{;},
  morecomment=[s]{/*}{*/},
  morestring=[b]",
  sensitive = true,
  classoffset=0,
  morecomment=[l]\#,
  morekeywords={
      service,
      file,
      package,
      if, 
      class,
      define,
      exec,
      include,
  },
  classoffset=1, keywordstyle=\ttfamily\color[rgb]{0,0,1},
  morekeywords={
      notify,
      require,
      subscribe,
      File,
      Package,
      mode,
      group,
      owner,
      provider,
      source,
      loc,
      command,
      creates,
      export,
      unless,
      replace,
      true,
      false,
  },
  alsoletter={\%},
  keywordsprefix={\%},
}
\lstdefinestyle{bugcode}{language=C++, basicstyle=\scriptsize\ttfamily,
  numbers=left, numberstyle=\tiny, numbersep=6pt, xleftmargin=11pt,
  aboveskip=0pt, belowskip=0pt, breaklines=false, keepspaces=true,
  backgroundcolor={}, escapeinside={(*@}{@*)}}
\usetikzlibrary{tikzmark,calc,positioning,arrows.meta}
\definecolor{unsafefill}{HTML}{FDECEC}
\definecolor{unsafedraw}{HTML}{F2B8B8}
\tikzset{unsafebox/.style={fill=unsafefill, draw=unsafedraw,
  line width=0.4pt, rounded corners=2pt}}
\newcounter{unsafebox}
\newlength{\unsafeleft}

\lstdefinelanguage{JavaScript}{
  morekeywords={const, let, var, new, function, return, get, set,
    true, false, null, undefined, typeof, this, require},
  sensitive=true,
  morecomment=[l]{//}, morecomment=[s]{/*}{*/},
  morestring=[b]', morestring=[b]", morestring=[b]`}
\lstdefinestyle{bugrepl}{language=JavaScript, basicstyle=\scriptsize\ttfamily,
  numbers=none, xleftmargin=0pt, aboveskip=0pt, belowskip=0pt,
  breaklines=false}
\newcommand{\promptfont}{\fontsize{7}{8.6}\selectfont\ttfamily}
\newlength{\promptch}
\newcommand{\promptsetup}{\promptfont\setlength{\promptch}{\fontcharwd\font`\x}%
  \raggedright\hyphenpenalty=10000 \exhyphenpenalty=10000 \parindent=0pt \parskip=0pt\relax
  \microtypesetup{protrusion=false}\promptquote}
\begingroup\lccode`\~=`\'
\lowercase{\endgroup\newcommand{\promptquote}{\catcode`\'=\active\def~{\textquotesingle}}}
\newtcolorbox{promptbox*}{enhanced, breakable=false, width=\linewidth,
  colback=white, colframe=black!35, boxrule=0.4pt, arc=1.5pt,
  boxsep=0pt, left=8pt, right=8pt, top=6pt, bottom=6pt,
  fontupper=\promptsetup}
\newcommand{\prompthead}[1]{\par\ifdim\prevdepth>-1000pt\addvspace{3.5pt}\fi
  {\bfseries #1}\par\nobreak}
\newcommand{\promptitem}[3][0]{\par\hangindent=2\promptch\hangafter=0
  \textbf{#2} #3\par}
\newcommand{\promptsubitem}[2]{\par\hangindent=4\promptch\hangafter=0
  \textbf{#1} #2\par}
\newcommand{\promptbullet}[1]{\par\hangindent=2\promptch\hangafter=1
  \makebox[2\promptch][l]{-}#1\par}
\newcommand{\elided}{[...]}
\newcommand{\replsep}{\par\vspace{3pt}%
  {\color{black!50}\rule{\linewidth}{0.4pt}}\par\vspace{1pt}}

\newcommand{\tool}{\textsc{Xtypecheck}}

\newcommand{\numNpmPackages}{667}
\newcommand{\numPyPIModules}{\nnum{1052}}
\newcommand{\numNpmAnalyzed}{662}
\newcommand{\numPyPIAnalyzed}{\nnum{1030}}

\newcommand{\empirical}[1]{#1}

\newcommand{\point}[1]{\noindent{\bf #1:} }

\newtcolorbox{promptbox}[1][]{
  colback=lightgray,
  colframe=black,
  arc=2mm,
  boxrule=0.4pt,
  left=4pt,right=4pt,top=4pt,bottom=4pt,
    fontupper=\tiny\normalfont, 
  width=\textwidth,
   enhanced jigsaw,
  #1
}

\newcommand{\grouprule}{\noalign{\vskip 1.5pt
  \begingroup\color{black!25}\hrule height 0.3pt\endgroup\vskip 1.5pt}}
\definecolor{groupshade}{gray}{0.93}

\providecommand{\llbracket}{[\![}\providecommand{\rrbracket}{]\!]}

\newcommand{\code}[1]{\texttt{#1}}
\newcommand{\kw}[1]{\mathsf{#1}}

\newcommand{\Any}{\mathsf{Any}}
\newcommand{\AnyObj}{\kw{Object}}

\newcommand{\None}{\mathsf{Failed}}
\newcommand{\tnum}{\mathsf{Num}}\newcommand{\tstr}{\mathsf{Str}}\newcommand{\tbool}{\mathsf{Bool}}\newcommand{\tbytes}{\mathsf{Bytes}}\newcommand{\AsciiStr}{\mathsf{AsciiStr}}
\newcommand{\tnull}{\mathsf{Null}}\newcommand{\tundef}{\mathsf{Undef}}

\newcommand{\exact}[1]{#1^{=}}
\newcommand{\sem}[1]{\llbracket #1 \rrbracket}
\newcommand{\vof}[1]{M(\texttt{#1})} 
\newcommand{\ischk}[2]{\mathsf{is}_{#1}(#2)}
\newcommand{\proj}[2]{\mathsf{use}_{#1}(#2)}

\newcommand{\wrong}{\mathbf{wrong}}
\newcommand{\RN}[1]{\textsc{#1}}
\newcommand{\figrule}{\par\vspace{2pt}\noindent{\color{black!60}\hrule height 0.4pt}\vspace{-4pt}}
\AtBeginDocument{}
\newcommand{\Let}[2]{\kw{let}\;#1=#2}
\newcommand{\Ifs}[3]{\kw{if}\;#1\;\kw{then}\;#2\;\kw{else}\;#3}

\newcommand{\Host}[2]{\kw{hop}_{#1}(#2)}
\newcommand{\Arg}[1]{\kw{arg}(#1)}
\newcommand{\Cfg}[2]{\langle #1,\,#2\rangle}
\newcommand{\env}{E} 
\newcommand{\cat}{\mathbin{+\!\!+}}
\newcommand{\sq}{\overline{s}}
\newcommand{\eseq}{\varepsilon}
\newcommand{\emp}{\mathsf{failed}}

\AtBeginDocument{\theoremstyle{acmdefinition}}
\newtheoremstyle{defbold}{\medskipamount}{\medskipamount}{\normalfont}{}{\bfseries}{.}{0.5em}{\thmname{#1}\thmnumber{ #2}\thmnote{ (#3)}}
\AtBeginDocument{\theoremstyle{defbold}\newtheorem{defn}[theorem]{Definition}}

\newcommand{\xln}[1]{\text{\textcolor{black!40}{\sffamily\fontsize{6}{7}\selectfont #1}}}

\newcommand{\fixbox}[1]{\tikz[baseline=(fb.base)]{\node[draw=chkdraw,dash pattern=on 1.6pt off 1.1pt,line width=0.5pt,fill=chkfill,rounded corners=1pt,inner sep=1.5pt,inner xsep=2pt,anchor=base](fb){$#1$};}}
\setminted{fontsize=\scriptsize, numbersep=6pt, xleftmargin=11pt, breaklines=false, highlightcolor=unsafefill}
\newcommand{\api}[1]{\textbf{#1}} 
\usetikzlibrary{positioning,arrows.meta}
\crefname{figure}{Figure}{Figures}\Crefname{figure}{Figure}{Figures}
\crefname{subfigure}{Figure}{Figures}\Crefname{subfigure}{Figure}{Figures}
\crefname{section}{Section}{Sections}\Crefname{section}{Section}{Sections}
\crefname{table}{Table}{Tables}\Crefname{table}{Table}{Tables}
\crefname{appendix}{Appendix}{Appendices}\Crefname{appendix}{Appendix}{Appendices}
\crefname{theorem}{Theorem}{Theorems}\Crefname{theorem}{Theorem}{Theorems}

\definecolor{chkfill}{HTML}{DCE9F7}
\definecolor{safefill}{HTML}{E1F0E1}
\definecolor{badfill}{HTML}{F7D3D3}
\definecolor{weakfill}{HTML}{FBE5C8}
\definecolor{argfill}{HTML}{E4DAF4} 
\definecolor{chkdraw}{HTML}{3A6EA5}
\definecolor{regdraw}{HTML}{8B5A2B} 
\colorlet{baddraw}{red!60!black}
\newcounter{hlbox}
\newlength{\hlpad}
\newlength{\hlgap}
\newcommand{\hlrad}{1.3pt}
\makeatletter
\newlength{\hl@tw}
\def\hl@dashed{d}
\newcommand{\hl@draw}[3]{
  \stepcounter{hlbox}\edef\hl@n{\thehlbox}%
  \iftikzmark{hl-c-\hl@n}{\def\hl@c{hl-c-\hl@n}}{\def\hl@c{hl-e-\hl@n}}%
  \tikz[remember picture,overlay]{%
    \path let \p1=(pic cs:hl-e-\hl@n), \p2=(pic cs:\hl@c) in
      \pgfextra{\xdef\hl@xe{\x1}\xdef\hl@ye{\y1}\xdef\hl@xc{\x2}};
    \pgfmathsetlengthmacro\hl@t{\ht\strutbox-0.5\hlgap}%
    \pgfmathsetlengthmacro\hl@b{\hl@ye-\dp\strutbox+0.5\hlgap}%
    \ifx\relax#1\relax
      \pgfmathsetlengthmacro\hl@r{max(\hl@xe,\hl@xc)+\hlpad}%
    \else
      \settowidth{\hl@tw}{\sitelab{#1}}\addtolength{\hl@tw}{2.6pt}%
      \pgfmathsetlengthmacro\hl@r{max(\hl@xc+\hlpad,\hl@xe+3.2pt+\hl@tw)}%
    \fi
    \fill[#3, rounded corners=\hlrad] (-\hlpad,\hl@b) rectangle (\hl@r,\hl@t);
    \ifx\relax#1\relax\else
      \begin{scope}
        \clip[rounded corners=\hlrad] (-\hlpad,\hl@b) rectangle (\hl@r,\hl@t);
        \fill[baddraw] ({\hl@r-\hl@tw},\hl@b) rectangle (\hl@r,{\hl@b+\baselineskip-\hlgap});
      \end{scope}
      \node[anchor=base, inner sep=0pt] at ({\hl@r-0.5\hl@tw},\hl@ye) {\sitelab{#1}};
    \fi
    \def\hl@arg{#2}\ifx\hl@arg\hl@dashed
      \draw[chkdraw, line width=0.5pt, dash pattern=on 1.6pt off 1.1pt,
        rounded corners=\hlrad-0.25pt]
        ({-\hlpad+0.25pt},{\hl@b+0.25pt}) rectangle ({\hl@r-0.25pt},{\hl@t-0.25pt});
    \fi}}
\newcommand{\hb}[2][]{\hl@draw{#1}{}{#2}}
\newcommand{\hbd}[2][]{\hl@draw{#1}{d}{#2}}
\makeatother
\newcommand{\he}{\tikzmark{hl-e-\thehlbox}}
\newcommand{\hc}{\tikzmark{hl-c-\thehlbox}}

\newcommand{\sitelab}[1]{{\scriptsize\color{white}$\mathsf{s}_{\mathsf{#1}}$}}
\newcommand{\fixl}{\llap{\textcolor{chkdraw}{\bfseries +}\hspace{1.2pt}}}

\colorlet{repldraw}{black!18}
\newtcolorbox{replbox}{colback=white, colframe=repldraw, boxrule=0.3pt,
  sharp corners, boxsep=0pt, left=2.5pt, right=2.5pt, top=2pt, bottom=2pt,
  grow to left by=2.5pt, before skip=4pt, after skip=0pt}

\newcommand{\replo}[1]{\textcolor{baddraw}{#1}}
\definecolor{kwcolor}{HTML}{5E3384}
\definecolor{strcolor}{HTML}{2E6B3A}
\makeatletter
\newcommand{\codepalette}{%
  \@namedef{PYG@tok@o}{}\@namedef{PYG@tok@m}{}\@namedef{PYG@tok@mi}{}\@namedef{PYG@tok@nf}{}%
  \@namedef{PYG@tok@k}{\def\PYG@tc####1{\textcolor{kwcolor}{####1}}}%
  \@namedef{PYG@tok@kt}{\def\PYG@tc####1{\textcolor{kwcolor}{####1}}}%
  \@namedef{PYG@tok@nb}{\def\PYG@tc####1{\textcolor{kwcolor}{####1}}}%
  \@namedef{PYG@tok@s}{\def\PYG@tc####1{\textcolor{strcolor}{####1}}}}
\makeatother
\newcommand{\Hs}[1]{H_{\mathrm{#1}}} 
\newcommand{\tor}{\vee} 

\usepackage{float}

%% file: figures/motivating.tex
\begin{figure}[t]
\renewcommand{\theFancyVerbLine}{\textcolor{black!40}{\sffamily\fontsize{6}{7}\selectfont\arabic{FancyVerbLine}}}%
\newcommand{\scriptlines}[1]{\renewcommand{\theFancyVerbLine}{\textcolor{black!40}{\sffamily\fontsize{6}{7}\selectfont
  \ifnum\value{FancyVerbLine}>#1 \else S\arabic{FancyVerbLine}\fi}}}%
\setminted{fontsize=\fontsize{7.2}{10.5}\selectfont, xleftmargin=13.6pt, numbersep=7.6pt, escapeinside=@@, formatcom=\codepalette}%
\setlength{\hlgap}{2.5pt}\renewcommand{\hlrad}{0pt}
\captionsetup[sub]{justification=centering,singlelinecheck=false,font=footnotesize,position=below,skip=3pt}
\begin{minipage}[t]{\dimexpr0.5\textwidth-6.4pt}
\vspace{0pt}
\begin{minipage}[t]{\linewidth}
\scriptlines{4}%
\begin{minted}[linenos]{text}
@\replp@ ({ createCanvas } = require('canvas'))
@\replp@ cv = createCanvas(10, 10)
@\replp@ cv.context = {}
@\replp@ cv.width = 20
@\replo{Segmentation fault}@
\end{minted}
\end{minipage}
\replsep
\begin{minipage}[t]{\linewidth}
\begin{minted}[linenos]{cpp}
void Canvas::resurface(Object This, ...) {
  Value ctx = This.@\api{Get}@("context");
  if (@\hb{weakfill}@!ctx.@\api{IsEmpty}@() && ctx.@\api{IsObject}@()@\he@) {
@\fixl@    @\hb{chkfill!60}@if (ctx.@\api{CheckTypeTag}@(&tags.Context2d)) {@\he@
      Context2d *c2d = @\hb{badfill}@Context2d::@\api{Unwrap}@(ctx)@\he@;
      cairo_t *prev = c2d->context();
      ...
@\fixl@    @\hb{chkfill!60}@}@\he@
  }
}
void Canvas::SetWidth(CallbackInfo& info) {
  ...
  resurface(info.This().As<Object>(), ...);
}
\end{minted}
\end{minipage}
\end{minipage}\hfill
\begin{minipage}[t]{\dimexpr0.5\textwidth-6.4pt}
\vspace{0pt}
\begin{minipage}[t]{\linewidth}
\scriptlines{7}%
\begin{minted}[linenos]{text}
@\replpy@ import codecs
@\replpy@ def noop_encode(s): return s
@\replpy@ codecs.register("noop", noop_encode)
@\replpy@ f = open("f", "w", encoding="noop")
@\replpy@ f.write("DELETE: NO™")
@\replpy@ f.close()
@\replpy@ print(open("f").read())
@\replo{DELETE}\textcolor{black!40}{\sbox0{: NO™}\rlap{\rule[0.6ex]{\wd0}{0.4pt}}\usebox0}@
\end{minted}
\end{minipage}
\replsep
\begin{minipage}[t]{\linewidth}
\begin{minted}[linenos]{c}
PyObject *write(textio *self, PyObject *text) {
  b = @\api{PyObject\_CallOneArg}@(encode, text);
  if (b == NULL) return NULL;
  if (@\hb{weakfill}@b != text && !@\api{PyBytes\_Check}@(b)@\he@)
    return TypeError;
@\fixl@  @\hb{chkfill!60}@if (b == text && !@\api{PyUnicode\_IS\_ASCII}@(b))@\hc@
@\fixl@    return TypeError;@\he@
  if (@\api{PyUnicode\_Check}@(b)) {
    n = @\api{PyUnicode\_GET\_LENGTH}@(b);
    memcpy(out, @\hb{badfill}\api{PyUnicode\_1BYTE\_DATA}@(b)@\he@, n);
  }
\end{minted}
\end{minipage}
\end{minipage}\par
\begin{minipage}[t]{\dimexpr0.5\textwidth-6.4pt}
\subcaption{Segmentation fault in Node.js canvas 3.2.3}\label{fig:motivating-a}
\end{minipage}\hfill
\begin{minipage}[t]{\dimexpr0.5\textwidth-6.4pt}
\subcaption{Silent output truncation in Python's built-in \code{io}}\label{fig:motivating-b}
\end{minipage}
\vspace{-8pt}
\caption{Two boundary type errors found by our approach together with their reproducers.
Red marks the uses that cause the errors, orange the incomplete type checks,
and blue the fixes.}
\label{fig:motivating}
\end{figure}

%% file: figures/fig-interaction.tex
\begin{figure}[t]
\centering
\colorlet{safedraw}{green!40!black}
\definecolor{kwcolor}{HTML}{5E3384}
\newcommand{\iapi}[1]{\textbf{#1}}
\newcommand{\ikwd}[1]{\textcolor{kwcolor}{#1}}
\newcommand{\ipr}[1]{\textcolor{black!40}{#1}}
\newcommand{\icodefont}{\ttfamily\fontsize{7.5}{10.5}\selectfont}
\newlength{\icw}
\settowidth{\icw}{\icodefont x}%
\scalebox{0.9}{%
\begin{tikzpicture}[x=1mm, y=1mm, line width=0.4pt,
  >={Latex[length=3.6pt]},
  code/.style={font=\icodefont, inner xsep=1.6pt, inner ysep=0.9pt,
    text height=5.4pt, text depth=1.7pt, anchor=base west, rounded corners=1.3pt},
  lno/.style={font=\sffamily\fontsize{6}{7}\selectfont, text=black!40,
    anchor=base east, inner sep=0pt},
  role/.style={font=\sffamily\bfseries\fontsize{6.5}{7.5}\selectfont, anchor=base east, inner sep=0pt},
  head/.style={font=\sffamily\bfseries\fontsize{8}{9}\selectfont, inner sep=0pt},
  note/.style={font=\sffamily\fontsize{6.5}{7.5}\selectfont, inner sep=0pt, align=left},
  arr/.style={->, line width=0.5pt, draw=#1},
  arr/.default=black,
]
\def\p{3.7}                 
\def\xr{15.2}               
\def\xn{17.8}               
\def\xc{18.8}               
\def\xa{74.4}\def\wa{8.0}   
\def\xh{85.6}               
\def\W{135.4}               
\def\xv{\xh+1.6}            
\def\wv{30.0}\def\wl{10.0}  
\def\hT{4.2}\def\hB{6.2}    
\def\xvc{24.7}              
\def\hc{3.9}                
\def\yC{-0.2}               
\def\dc{0.65}               
\pgfmathsetmacro\yA{\yC-3*\hc-3.6}    
\newcommand{\Yp}[1]{\fpeval{\yA-(#1-1)*\p}}   
\newcommand{\Y}[1]{\ifcase#1 0\or\Yp{1}\or\Yp{2}\or\Yp{3}\or\Yp{4}\or\Yp{5}\or\yFive\or\ySeven\or\ySix\or\fpeval{\ySix-\p}\or\fpeval{\ySix-2*\p}\fi}
\newcommand{\C}[1]{\fpeval{\Y{#1}+\dc}}  
\tikzset{small/.style={note, font=\sffamily\fontsize{5.5}{6.5}\selectfont}}
\newcommand{\hv}[6]{%
  \filldraw[draw=black!60, fill=#5] (\xv,#2) rectangle (\xv+\wv+\wl,#2-\hT);
  \filldraw[draw=black!60, fill=#6] (\xv,#2-\hT) rectangle (\xv+\wv+\wl,#2-\hT-\hB);
  \node[font=\sffamily\bfseries\fontsize{7}{8}\selectfont, inner sep=0pt, anchor=west] at (\xv+2.0,#2-\hT/2) {tag:};
  \node[font=\ttfamily\fontsize{7}{8}\selectfont, inner sep=0pt] at (\xv+\xvc,#2-\hT/2) {#3};
  \node[font=\sffamily\bfseries\fontsize{7}{8}\selectfont, inner sep=0pt, anchor=west] at (\xv+2.0,#2-\hT-\hB/2) {layout:};
  \node[font=\sffamily\fontsize{7}{8}\selectfont, inner sep=0pt] at (\xv+\xvc,#2-\hT-\hB/2) {raw bits};
  \coordinate (#1t) at (\xv,#2-\hT/2);
  \coordinate (#1b) at (\xv,#2-\hT-\hB/2);}
\pgfmathsetmacro\yVa{\fpeval{\Yp{3}+\dc}+\hT/2}  
\pgfmathsetmacro\yVb{\yVa-\hT-\hB-9.6}            
\pgfmathsetmacro\ySix{\yVb-\hT-\hB/2-\dc+1.3}         
\pgfmathsetmacro\yFive{\Yp{5}-(\Yp{5}-\ySix)/3}    
\pgfmathsetmacro\ySeven{\Yp{5}-2*(\Yp{5}-\ySix)/3} 
\node[head, anchor=base west] at (\xc,-0.4) {Native code};
\newcommand{\cl}[4]{
  \node[code, fill=#3] (L#1) at ($(\xc,\Y{#1})+(#2\icw,0)$) {#4};
  \node[lno] at (\xn,\Y{#1}) {#1};}
\cl{1}{0}{none}{Value Total(Args info) \{}
\cl{2}{2}{none}{Value a = info[0], f = info[1];}
\cl{3}{2}{chkfill}{\ikwd{if} (!a->\iapi{IsArray}()) \ikwd{return};}
\cl{4}{2}{safefill}{\ikwd{int} n = a->\iapi{Length}();}
\cl{5}{2}{chkfill}{\ikwd{if} (!f->\iapi{IsFunction}()) \ikwd{return};}
\cl{6}{2}{safefill}{Value b = f->\iapi{Call}();}
\cl{7}{2}{weakfill}{\ikwd{if} (b.\iapi{IsEmpty}()) \ikwd{return};}
\cl{8}{2}{badfill}{\ikwd{double} d = b.\iapi{As}<Number>()->\iapi{Value}();}
\cl{9}{2}{none}{\ikwd{return} ...;}
\cl{10}{0}{none}{\}}
\foreach \l/\r in {1/native entry point, 2/argument read, 3/test, 4/use, 5/test, 7/test, 8/use, 9/return}{
  \node[role] at (\xr,\Y{\l}) {\r};
}
\node[role, anchor=east, align=right] at (\xr,\C{6}) {use\\host operation};
\def\xg{55.0}\def\wg{13.0}
\tikzset{acell/.style={draw=black!60, fill=white, minimum width=\wg mm, minimum height=\hc mm,
  inner sep=0pt, anchor=north west, font=\ttfamily\fontsize{7}{8}\selectfont}}
\coordinate (infoc) at ($(\xc,\yC)+(19\icw,0)$);   
\pgfmathsetmacro\yHt{-2.2}\pgfmathsetmacro\yHb{\yVb-\hT-\hB-1.8}
\draw[black!55, rounded corners=1.5pt] (\xh,\yHt) rectangle (\W,\yHb);
\node[head, anchor=base] at ({(\xh+\W)/2},-0.4) {Host};
\pgfmathsetmacro\yS{\yC-2.5*\hc-\dc}  
\draw[repldraw, fill=white] (\xh+1.6,\yS+\p+3.1) rectangle (\W-1.6,\yS-\p-2.0);
\node[font=\sffamily\bfseries\fontsize{7}{8}\selectfont, inner sep=0pt, anchor=south east] at (\W-2.4,\yS-\p-1.5) {Script};
\node[code, inner xsep=0pt] at (\xh+2.6,\yS+\p) {\ipr{>} \ikwd{const} ext = require(\textquotesingle ext\textquotesingle)};
\node[code, inner xsep=0pt] (S1) at (\xh+2.6,\yS) {\ipr{>} ext.total([3, 4], () => \textquotedbl x\textquotedbl)};
\node[code, inner xsep=0pt, text=baddraw] at ($(\xh+2.6,\yS-\p)+(2\icw,0)$) {6.9e-310 (garbage)};
\hv{va}{\yVa}{Array}{the argument}{chkfill}{safefill}
\hv{vb}{\yVb}{String}{its result}{white}{badfill}
\fill[black!8] (\xa,-5.4) rectangle (\xa+\wa,\yHb);
\node[head, anchor=north, align=center] at (\xa+\wa/2,2.0) {binding\\API};
\draw[arr] ([xshift=-0.3mm]S1.west) -| node[pos=0.3, above=0.3mm, font=\sffamily\fontsize{7}{8}\selectfont, inner sep=0pt] {\textit{anything}} (infoc |- L1.north);  
\draw[arr=chkdraw] (L3.east) -- (L3.east -| \xv,0);              
\draw[arr=safedraw] (L4.east) -- (L4.east -| \xv,0);
\draw[arr=chkdraw, dash pattern=on 1.6pt off 1.1pt] (L5.east) -- (L5.east -| \xa,0);  
\draw[arr] ([yshift=0.9mm]L6.east) -- node[pos=0.42, above=0.3mm, font=\sffamily\fontsize{7}{8}\selectfont, inner sep=0pt] {runs \textbf{host code}} ([yshift=0.9mm]L6.east -| \xh,0);
\draw[arr] ([yshift=-0.9mm]L6.east -| \xh,0) -- node[pos=0.58, below=0.3mm, font=\sffamily\fontsize{7}{8}\selectfont, inner sep=0pt] {\textit{anything} or $\emp$} ([yshift=-0.9mm]L6.east);       
\draw[arr=baddraw] (L8.east) -- (L8.east -| \xv,0);    
\end{tikzpicture}}
\vspace{-5pt}
\caption{
A~\textit{script} (JavaScript),
its~\textit{host} (V8),
and~\textit{native code} interacting
through a binding API.
}
\label{fig:interaction}
\vspace{-6pt}
\end{figure}

%% file: figures/fig-taginf.tex
\begin{tikzpicture}[x=1cm, y=1cm,
  vf/.style={draw=black!60, rounded corners=1.5pt, font=\normalsize, inner sep=2.2pt, minimum height=5.2mm, fill=white},
  vftest/.style={vf, fill=weakfill},
  vfbad/.style={vf, fill=badfill},
  vfreg/.style={-{Stealth[length=1.5mm]}, line width=0.4pt},
  vflab/.style={font=\small, fill=white, inner sep=1pt},
  tf/.style={font=\ttfamily\footnotesize, text=black!65, inner sep=0.5pt},
  jn/.style={circle, draw=black!60, fill=white, inner sep=0pt, minimum size=3pt},
  vfret/.style={vf, font=\small, inner sep=1.8pt, minimum height=4.2mm},
]
\def\xe{-1.75} 
\node[vftest] (d3)  at (0,0)      {$\ischk{\None}{b}$};
\node[vf]     (st)  at (0,-1.0)  {$\star$};
\node[vftest] (db)  at (0,-2.0)   {$\ischk{\tbytes}{b}$};
\node[jn]     (J)   at (0,-2.95)   {};
\node[vf, font=\small, inner sep=1.6pt, minimum height=3.8mm] (eps) at (\xe,-2.0) {$\eseq$};
\node[vftest] (d6)  at (0,-3.85)  {$\ischk{\tstr}{b}$};
\node[vf]     (u7)  at (0,-4.85)   {$\proj{\tstr}{b}$};
\node[vfbad]  (u8)  at (0,-5.85)  {$\proj{\AsciiStr}{b}$};
\def\xr{2.15}
\node[vfret] (r3)  at (\xr,0)      {$\kw{return}$};
\node[vfret] (r4)  at (\xr,-2.0)   {$\kw{return}$};
\node[vfret] (r6)  at (\xr,-3.85)  {$\kw{return}$};
\draw[vfreg] (d3) -- (r3);
\node[tf, anchor=south west] at ($(d3.east)+(0.06,0.04)$) {T};
\draw[vfreg] (db) -- (r4);
\node[tf, anchor=south west] at ($(db.east)+(0.06,0.04)$) {F};
\draw[vfreg] (d6) -- (r6);
\node[tf, anchor=south west] at ($(d6.east)+(0.06,0.04)$) {F};
\draw[vfreg] (0,0.7) -- node[vflab, right=3pt] {$\Gamma(b)=\Any$} (d3);
\draw[vfreg] (d3) -- node[vflab, right=3pt] {$\neg\None$} (st);
\node[tf, anchor=north east] at ($(d3.south)+(-0.06,-0.04)$) {F};
\draw[vfreg] (st) -- (db);
\node[tf, anchor=north east] at ($(st.south)+(-0.06,-0.04)$) {T};
\draw[vfreg] (st) -- node[vflab, above=1.5pt, pos=0.58] {$\neg\None$} (\xe,-1.0) -- (eps);
\draw[vfreg] (eps) -- (\xe,-2.95) -- (J);
\node[tf, anchor=south east] at ($(st.west)+(-0.06,0.04)$) {F};
\draw[vfreg] (db) -- node[vflab, right=3pt] {$\neg\None\wedge\tbytes$} (J);
\node[tf, anchor=north east] at ($(db.south)+(-0.06,-0.04)$) {T};
\draw[vfreg] (J) -- node[vflab, right=3pt] {$\neg\None$} (d6);
\draw[vfreg] (d6) -- node[vflab, right=3pt] {$\neg\None\wedge\tstr$} (u7);
\node[tf, anchor=north east] at ($(d6.south)+(-0.06,-0.04)$) {T};
\draw[vfreg] (u7) -- node[vflab, right=3pt] {$\neg\None\wedge\tstr$} (u8);
\end{tikzpicture}